\documentclass[amsmath,amssymb,10pt]{revtex4}
\usepackage{amsfonts}
\usepackage{bm}
\usepackage{amsmath}
\begin{document}

\title{Free energy of a Lovelock holographic superconductor}

\author{Ligeia Aranguiz}
\email{ligeia.aranguiz@postgrado.usm.cl}
\affiliation{Instituto de F\'{i}sica, Pontificia Universidad Cat\'{o}lica de
Valpara\'{i}so, \\ Casilla 4059, Valpara\'{i}so, Chile}
\affiliation{Universidad T\'{e}cnica Federico Santa Mar\'{\i}a, Casilla
110-V, Valpara\'{\i}so, Chile}

\author{Olivera Miskovic}
\email{olivera.miskovic@ucv.cl}
\affiliation{Instituto de F\'{i}sica, Pontificia Universidad Cat\'{o}lica de
Valpara\'{i}so, \\ Casilla 4059, Valpara\'{i}so, Chile }

\begin{abstract}
We study thermodynamics of black hole solutions in Lanczos-Lovelock AdS gravity in $d+1$
dimensions coupled to nonlinear electrodynamics and a St\"{u}ckelberg scalar
field. This class of theories is used in
the context of gauge/gravity duality to describe a high-temperature
superconductor in $d$ dimensions. Larger number of coupling constants in the
gravitational side is necessary to widen a domain of validity of physical quantities in a dual QFT.
We regularize the gravitational action and
find the finite conserved quantities for a planar black hole with scalar
hair. Then we derive the quantum statistical relation in the Euclidean
sector of the theory, and obtain the exact formula for the free energy of
the superconductor in the holographic quantum field theory. Our result is
analytic and it includes the effects of backreaction of the
gravitational field. We further discuss on how this formula could be used to
analyze second order phase transitions through the discontinuities of the
free energy, in order to classify holographic superconductors in terms of the
parameters in the theory.
\end{abstract}

\maketitle

\section{Introduction}

The proposal of Maldacena about an equivalence between the anti-de Sitter
(AdS) gravity and a conformal field theory (CFT) in a fewer spacetime
dimension \cite{AdS/CFT} has brought to many successful
applications of this correspondence to strongly coupled quantum systems.
This conjectured holographic-type duality between two theories is still the
only approach to field theories where strong coupling results are calculated
exactly from its gravitational weakly coupled dual system.

We are interested in studying thermal phase transitions in quantum field
theories in the context of AdS/CFT correspondence. The temperature in a
holographic quantum field theory (QFT) is identified with the Hawking
temperature of the black hole in AdS gravity which arises from quantum
effects near the event horizon. A phase transition occurs in a theory when
its effective potential changes as a consequence of temperature variations
and leads to vacuum instabilities. We assume that the temperatures are high
enough in order to favor transitions due to temperature fluctuations, and
not the ones that happen because of quantum fluctuations. In particular, we
shall focus\textbf{\ }on high-temperature superconductors whose critical
temperature, $T_c$, is above the limit of applicability of the
Bardeen-Cooper-Schrieffer (BCS) theory, and use gauge/gravity duality\textbf{%
\ }to describe\textbf{\ }phase transitions as a fundamental phenomenon and
not an effective one, as in the BCS model.

Typical features of a superconducting phase in a superconductor are related
to its response to electric and magnetic fields, such as the infinite
conductivity (i.e., superconductivity) and the expulsion of magnetic field
from it (the Meissner effect). Thus, in order to obtain a holographic
superconductor in $d$ dimensions, the dual AdS$_{d+1}$ space has to contain
an electrically charged black hole coupled to charged matter, for example, a
complex scalar field \cite{Hartnoll:2008vx}. Then the electromagnetic field
becomes a source for an electromagnetic current in the boundary QFT and the
scalar field couples to an order parameter. In addition, in the gravity side
a charged black hole will develop a scalar hair as the Hawking temperature
increases, which will correspond to a phase transition of a superconductor
in the QFT side. The holographic procedure enables to compute dynamical
transport properties of a such superconductor directly from the
gravitational dynamics.

Second order phase transitions are present in a holographic QFT dual to
General Relativity minimally coupled to the Maxwell field and one complex
scalar. Even in this simplest case, in order to obtain information about a
holographic superconductor, one faces the technical problem of solving
nonlinear field equations of matter interacting with gravity in AdS space.
The easiest way to circumvent this problem is to take a probe limit, that
is, to study the dynamics of the matter fields in the black hole background,
i.e., neglecting the gravitational backreaction. Then one might use
numerical methods to solve the equations of motion. For example, in Ref.\cite%
{Hartnoll:2008kx}, the authors study electric and magnetic fields in a
holographic St\"{u}ckelberg superconductor\ \cite{Stuckelberg} with a
minimally coupled scalar field in a four-dimensional planar AdS black hole
background. They calculate the critical exponent, superconductivity, the
energy gap, free energy and specific heat in this theory using numerical
methods. Similar problem in a background of a planar black hole with a
monopole and a multiplet of complex scalar fields in four dimensions was
discussed in Ref.\cite{Chen:2009vz}. In $d$ dimensions, a more general
system has been analyzed in Refs.\cite{Franco:2009yz}, where a
number of physical quantities was obtained for the St\"{u}ckelberg
holographic superconductor with the scalar potential $\Psi ^{n}$, finding
first order ($n>2$) and second order ($n=2$) phase transitions. A linear
combination of the potentials with different $n$ was also discussed. In the
same context, but adding Weyl corrections, it was found in Ref.\cite%
{Ma:2011zze} that the critical exponent does not depend on the Weyl
parameter, that just confirms its universal value $1/2$ that does not depend
on the details of the dynamics of the system. It is worthwhile mentioning
that the holographic models of high-temperature superconductivity also
successfully explain a large ratio of the energy gap to critical temperature
compared to conventional superconductors \cite%
{Hartnoll:2008vx,Hartnoll:2008kx}.

A semi-analytic approach to phase transitions in four dimensions is
beautifully described by Gubser \cite{Gubser:2008px} as a
symmetry breaking effect, where a superconductor is formed near the black
hole horizon. Different types of superconductors (s-wave \cite{Gangopadhyay:2012am}
and p-wave \cite{Momeni:2012ab}, whose order parameters are the scalar and the
vector, respectively) are explained using Abelian and non-Abelian gauge
fields. Analytic calculations in the asymptotic region near the phase
transition point are performed in five-dimensional AdS gravity in \cite%
{Herzog:2010vz} for the St\"{u}ckelberg equations of motion that are solved
in the probe limit. The backreaction of the hairy charged black hole in four
dimensions has been included in Ref.\cite{Ge:2011cw}, where a semi-analytic
solution is obtained by matching smoothly the near-horizon series with the
asymptotic solution at the matching point located between the horizon and
the infinity.

In order to explain different types of holographic high-$T_{c}$\
superconductors discovered in nature, which do not fit any theoretical description,
one needs a broad class of different CFTs coming from AdS
gravities in the bulk. A massive scalar field in AdS space can give as many
unitary CFTs on the boundary as there are possible boundary conditions, but
in general there are just few ones (for example Dirichlet, Neumann and mixed
boundary conditions) \cite{Klebanov:1999tb}. This is not enough to explain a
variety of properties of these superconductors. Thus, in higher dimensions,
it is natural to consider a generalization of General Relativity in AdS
space, that is, higher-curvature Lanczos-Lovelock AdS gravities \cite%
{Lovelock} that depend on a family of coupling constants and still
possess second order field equations in the metric field.
Presence of these couplings can change physical
properties of holographic quantities, such as critical
temperature and transport coefficients ratios. In recent work on
holographic superconductors with electromagnetic and charged scalar fields
that deal with nonlinear gravitational effects, Einstein-Gauss-Bonnet (EGB)
AdS gravity quadratic in curvature, was studied in an arbitrary dimension
\cite{Jing:2010cx} in the probe limit and using the numerical
methods. A typical result is that the Gauss-Bonnet (GB) coupling decreases the critical
temperature of the superconductor and thus makes the condensation harder
\cite{Pan:2011ah,Gregory:2010yr}. In these references,
an analytical approach to the condensation in a
holographic dual to EGB gravity that includes an effect of backreaction of
black holes in five dimensions was also discussed. Again, inclusion of the
backreaction of the gravitational field makes the condensation harder in
these theories. Interestingly, in Ref.\cite{Barclay:2010up} it was
found that for very strong GB couplings, the critical temperature begins to
increase. Another technique to analytically study the critical phenomena,
based on the variational method for the Sturm-Liouville operator, was used
in Ref.\cite{Siopsis:2010uq}. Different aspects of a holographic
superconductor with the GB term were also analyzed in Refs.\cite%
{Ge:2011cw,Jing:2010cx,Gregory:2010yr,Barclay:2010up,Barclay:2010nm}%
, showing that the higher-order curvature corrections can modify the
universal ratio of the gap in the frequency-dependent conductivity to
critical temperature ($\omega _{g}/T_{c}=8$). Similarly, in dual holographic
hydrodynamics, the universal ratio
bound  of the shear viscosity to entropy density, $\eta /s\leq 1/4\pi $ \cite{Buchel:2003tz},
can be changed by higher-order curvature terms  (see, e.g. Ref.\cite{Ge:2009ac}).
Another way to modify the ratio $\omega _{g}/T_{c}$ is in dilaton extensions of holographic superconductors \cite{Salvio:2012at}.

On the other hand, gauge theories which are described by nonlinear actions
for Abelian or non-Abelian connections have also become standard in the
context of Superstring Theory and provide richer physics in holographic
theories. For example, introduction of the Born-Infeld electrodynamics (in
presence of EGB AdS black holes) affects the formation of the scalar hair
since it changes the transition point of the phase transition from the
second order to the first order \cite{Jing:2010cx}. Also, power-Maxwell
electromagnetic field in the background of the Schwarzschild AdS black hole
makes the scalar hair harder to condensate for larger power parameter
\cite{Jing:2011vz}. In both cases the critical exponent of the system
remains the same as in the Landau-Ginzburg mean field theory, that is $1/2$,
and the same happens for a large class of nonlinear electrodynamics models.
Inclusion of a dynamical electromagnetic
field in holographic superconductors was discussed in Ref.\cite{Domenech:2010nf},
that was crucial to obtain properties such as the Meissner effect. Notice that, although
the magnetic field makes the condensate harder to form, a negative GB coupling enhances the condensation
when the field is not too strong \cite{Ge:2010aa}.

Newer results on nonlinear effects in holographic superconductors can be
found in Refs. \cite{Pan:2012jf}. For more on holographic superconductors,
see comprehensive lectures in, e.g.\cite{Hartnoll:2009sz}.

In our approach, we are interested in analytic study of the effects of
higher-order gravitational terms in AdS gravity and nonlinear
electrodynamics on phase transitions in high-$T_{c}$\ holographic
superconductors of the St\"{u}ckelberg type, that includes backreaction of
the black holes. The goal is to find an exact formula for the free energy of
a superconductor in an arbitrary dimension $d$, that is UV finite and it
depends on at most on $[d/2]$ gravitational coupling parameters of the
Lanczos-Lovelock action and two arbitrary functions stemming from the NED
and St\"{u}ckelberg actions. In the course of this, we shall renormalize the
IR sector of AdS$_{d+1}$ gravity, that corresponds to a UV renormalization
of a holographic quantum effective action. As a\ result, we shall obtain
that the free energy of a system at finite $T$ satisfies the Quantum
Statistical Relation. Since the free energy and the corresponding
statistical partition function contain all thermodynamic information about
the holographic quantum system, this formula will open a possibility to
analyze the local and global minima of the thermodynamic potential and, in
that way, detect all possible phase transitions in the theory depending on
the values of coupling constants, similarly as in the Landau-Ginzburg
description of superconductivity.

In this paper, we shall focus on the first part of the above problem, and
only discuss about the second part, that is work in progress.

\section{Lanczos-Lovelock AdS gravity and the equations of motion}

The Lanczos-Lovelock (LL) gravity \cite{Lovelock} in $D=d+1$
dimensions is described by an action polynomial in the Riemann curvature in
a such way that its equations of motion still keep properties of General
Relativity and give rise to at most second order field equations in the
metric. This happens because the $p$-th term of the LL polynomial, $L_{p}$,
is a dimensional continuation of the Euler density in $2p$ dimensions,
\begin{eqnarray}
I_{\text{LL}} &=&\frac{1}{2\kappa ^{2}}\int d^{d+1}x\sqrt{-g}%
\sum_{p=0}^{[d/2]}\alpha _{p}L_{p}\,,  \notag \\
L_{p} &=&\frac{1}{2^{p}}\,\delta _{\nu _{1}\cdots \nu _{2p}}^{\mu _{1}\cdots
\mu _{2p}}\,R_{\mu _{1}\mu _{2}}^{\nu _{1}\nu _{2}}\cdots R_{\mu _{2p-1}\mu
_{2p}}^{\nu _{2p-1}\nu _{2p}}\,,  \label{LL action}
\end{eqnarray}%
where $\delta _{\nu _{1}\cdots \nu _{2p}}^{\mu _{1}\cdots \mu _{2p}}=\delta
_{\nu _{1}}^{\mu _{1}}\delta _{\nu _{2}}^{\mu _{2}}\cdots \delta _{\nu
2_{p}}^{\mu _{2p}}+\cdots $ denotes the completely antisymmetric product of $2p$
Kronecker's deltas. In our notation, the metric field $g_{\mu \nu }$ is
mostly positive and the Riemann curvature reads $R_{\;\;\nu \alpha \beta
}^{\mu }=\partial _{\alpha }\Gamma _{\nu \beta }^{\mu }-\partial _{\beta
}\Gamma _{\nu \alpha }^{\mu }+\Gamma _{\lambda \alpha }^{\mu }\Gamma _{\nu
\beta }^{\lambda }-\Gamma _{\lambda \beta }^{\mu }\Gamma _{\nu \alpha
}^{\lambda }$.

The last non-vanishing term in the sum, $L_{d+1}$, is the Euler topological
invariant that does not contribute to the dynamics, so it has not been
included in the series. The terms with $L_{2p>d+1}$ are identically
vanishing. The strength of gravitational interaction is determined by the
Newton's constant $G_{N}=\kappa ^{2}/8\pi $. The gravitational part of the
theory depends on a set of the coupling constants, $\alpha _{p}$, of
dimension [length]$^{2p-2}$. The first term in the LL
polynomial is constant, $L_{0}=1$, so that $\alpha _{0}=-2\Lambda $ is
related to the cosmological constant, that we shall assume to be negative, $%
\Lambda =-d\left( d-1\right) /2\ell ^{2}$. Here, $\ell $ is the AdS radius.
The linear term in the curvature is the Einstein-Hilbert term $L_{1}=R$, normalized as $\alpha _{1}=1$.
Other terms can be seen as higher-order curvature
corrections to General Relativity. The simplest, quadratic correction, is given by the
GB term $L_{2}=R^{2}-4R_{\mu \nu }R^{\mu \nu }+R_{\mu \nu \lambda
\sigma }R^{\mu \nu \lambda \sigma }$ with the coupling $\alpha _{2}=\alpha $%
. Even though we assume that the constants $\alpha _{p}$ are arbitrary for $%
p\geq 2$, from the point of view of the AdS/CFT correspondence, there are
restrictions on their values related to preserving of the causality in the
boundary of asymptotically AdS spacetime \cite{deBoer:2009gx,Camanho:2009hu}.

The metric is coupled to the Abelian gauge field $A_{\mu }(x)$ with the
associated field strength $F_{\mu \nu }=\partial _{\mu }A_{\nu }-\partial
_{\nu }A_{\mu }$ through the quadratic invariant $F^{2}=g^{\mu \alpha
}g^{\nu \beta }F_{\mu \nu }F_{\alpha \beta }$. In order to include
non-linear effects, we choose the electromagnetic field described by
Nonlinear Electrodynamics (NED) whose Lagrangian density is an arbitrary
function in the invariant $F^{2}$,
\begin{equation}
I_{\text{NED}}=\frac{1}{2\kappa ^{2}}\int d^{d+1}x\,\sqrt{-g}\,\mathcal{L}%
(F^{2})\,.
\end{equation}%
The gravitational and NED fields are coupled to a complex scalar field $\hat{%
\Psi}=\Psi e^{ip}$, where non-linear effects are introduced through a
non-minimal coupling of the St\"{u}ckelberg action \cite{Stuckelberg},%
\begin{equation}
I_{\text{S}}=\frac{1}{2\kappa ^{2}}\int d^{d+1}x\,\sqrt{-g}\,\left[ -\frac{1%
}{2}\,(\partial \Psi )^{2}-\frac{1}{2}\,m^{2}\Psi ^{2}-\frac{1}{2}\,\mathcal{%
F}(\Psi )\left( \partial p-A\right) ^{2}\right] \,.
\end{equation}%
Here, $\Psi (x)$ and $p(x)$ are real scalar fields, and $\mathcal{F}(\Psi )$
is an arbitrary real function that satisfies $\mathcal{F}(0)=0$ and $%
\mathcal{F}(\Psi )\geq 0$ in order to ensure positivity of the kinetic term
for $p$. The minimal coupling between the scalar and EM fields is recovered
by choosing the interaction as $\mathcal{F}_{\text{minimal}}(\Psi )=\Psi
^{2} $.

The total bulk action,
\begin{equation}
I_{0}=I_{\text{LL}}[g]+I_{\text{NED}}[g,A]+I_{\text{S}}[g,A,\Psi ,p]\,,
\label{I0}
\end{equation}%
depends on a set of constants in the gravity part, $\kappa $, $\Lambda $ and
$\alpha _{p}$, and two arbitrary functions $\mathcal{L}(F^{2})$ and $%
\mathcal{F}(\Psi )$ determining completely the matter couplings.

Our goal is to understand for which gravitational parameter range and what
interaction with the matter (within a chosen class of theories) it is
possible to have a phase transition of second order in a holographically
dual QFT.

Provided the boundary terms have been added to the bulk action, that we
shall discuss later in detail, the action reaches an extremum for the
following equations of motion,%
\begin{eqnarray}
\delta g^{\mu \nu } &:&\quad -\sum_{p=0}^{[d/2]}\frac{\alpha _{p}}{2^{p+1}}%
\,g_{\nu \lambda }\delta _{\mu \mu _{1}\cdots \mu _{2p}}^{\lambda \nu
_{1}\cdots \nu _{2p}}\,R_{\nu _{1}\nu _{2}}^{\mu _{1}\mu _{2}}\cdots R_{\nu
_{2p-1}\nu _{2p}}^{\mu _{2p-1}\mu _{2p}}=T_{\mu \nu }\,,  \notag \\
\delta A_{\mu } &:&\quad \nabla _{\nu }\left( 4F^{\mu \nu }\dfrac{d\mathcal{L%
}}{dF^{2}}\right) =-\mathcal{F}(\Psi )\left( \nabla ^{\mu }p-A^{\mu }\right)
\,,  \notag \\
\delta \Psi &:&\quad \left( \nabla ^{2}-m^{2}\right) \Psi =\dfrac{1}{2}\,%
\dfrac{d\mathcal{F}}{d\Psi }\,\left( \nabla p-A\right) ^{2}\,,  \notag \\
\delta p &:&\quad \nabla _{\mu }\left[ \rule{0pt}{15pt}\mathcal{F}(\Psi
)\left( \nabla ^{\mu }p-A^{\mu }\right) \right] =0\,,
\label{covariant LL eom}
\end{eqnarray}%
where $\nabla _{\mu }$ is covariant derivative with respect to the affine
connection $\Gamma _{\nu \lambda }^{\mu }$. Symmetric energy-momentum tensor
for the matter fields, conveniently normalized as $T_{\mu \nu }=-\frac{%
4\kappa ^{2}}{\sqrt{-g}}\frac{\delta (I_{\text{NED}}+I_{\text{S}})}{\delta
g^{\mu \nu }}$, has the form%
\begin{eqnarray}
T_{\mu \nu } &=&\frac{1}{2}\,g_{\mu \nu }\,\mathcal{L}+\frac{d\mathcal{L}}{%
dF^{2}}\,2F_{\mu \lambda }F_{\ \nu }^{\lambda }-\frac{1}{4}\,g_{\mu \nu }%
\left[ (\partial \Psi )^{2}+m^{2}\Psi ^{2}+\mathcal{F}(\Psi )\left( \partial
p-A\right) ^{2}\right]  \notag \\
&&+\frac{1}{2}\,\partial _{\mu }\Psi \partial _{\nu }\Psi +\frac{1}{2}\,%
\mathcal{F}(\Psi )\,\left( \partial _{\mu }p-A_{\mu }\right) \left( \partial
_{\nu }p-A_{\nu }\right) \,.  \label{T}
\end{eqnarray}%
In order to ensure a non-negative energy density of the matter, we impose
the weak energy condition to the energy-momentum tensor, $w=-T_{\mu \nu
}\,u^{\mu }u^{\nu }\geq 0$, for a timelike unit vector $u^{\mu }$.

The system of equations (\ref{covariant LL eom}) can be simplified by
noticing that the last equation is not independent from the others and can
be consistently eliminated by fixing the U(1) gauge symmetry, $%
p(x)\rightarrow p(x)+\alpha (x)$. From now on, we set $p(x)=0$.

In order to have an asymptotically AdS spacetime, we assume that
there exists the AdS vacuum ($T_{\mu \nu }=0$) with constant curvature globally, $R_{\mu
_{1}\mu _{2}}^{\nu _{1}\nu _{2}}=-\frac{1}{\ell _{\text{eff}}^{2}}\,\delta
_{\mu _{1}\mu _{2}}^{\nu _{1}\nu _{2}}$, with an effective AdS radius, $%
\ell _{\text{eff}}$. Plugging
in this condition in the gravitational equation (\ref{covariant LL eom})
gives rise to a polynomial
\begin{equation}
0=\sum_{p=0}^{[d/2]}\frac{\alpha _{p}}{\left( d-2p\right) !}\,\left( -\ell _{%
\text{eff}}^{-2}\right) ^{p}\,,  \label{effective LL}
\end{equation}%
that has at most $[d/2]$ different roots $1/\ell _{\text{eff}}^{2}\,$ for a
given set of the coefficients $\{\alpha _{p}\}$.

Now we want to describe a charged AdS black hole solution to the
equations (\ref{covariant LL eom}). It is known that black holes exist
in pure LL AdS gravity \cite{Cai:2001dz,Garraffo:2008hu} and in LL AdS gravity coupled
to NED, in particular in Born-Infeld electrodynamics \cite{Aiello:2004rz}.
Their thermodynamics has also been studied \cite{Myers:1988ze}. For a
recent review on Lovelock gravities, see e.g. Ref. \cite{Padmanabhan:2013xyr},
and in the context of holography Ref.\cite{Edelstein:2013sna}.

\section{Charged planar black hole in Einstein-Gauss-Bonnet gravity}

We start from the EGB AdS action, the simplest LL gravity
different than General Relativity defined in $D\geq 5$ dimensions, that
includes a\ quadratic correction in the spacetime curvature given by the
GB term,
\begin{equation}
I_{\text{EGB}}=\frac{1}{2\kappa ^{2}}\int d^{d+1}x\,\sqrt{-g}\,\left[
R-2\Lambda +\alpha \,\left( R^{2}-4R_{\mu \nu }R^{\mu \nu }+R_{\mu \nu
\lambda \sigma }R^{\mu \nu \lambda \sigma }\right) \right] \,.
\end{equation}%
In the framework of String Theory, the GB term arises in the low-energy
limit and the constant $\alpha $ is positive. In our case, we are rather
concerned about restrictions imposed in a dual QFT, se we keep $\alpha $ an
arbitrary real constant. The gravitational equations of motion in (\ref%
{covariant LL eom}) becomes%
\begin{equation}
G_{\nu }^{\mu }+H_{\nu }^{\mu }=T_{\nu }^{\mu }\,,  \label{grav}
\end{equation}%
where we have introduced the Einstein tensor with the cosmological term,
\begin{equation}
G_{\nu }^{\mu }=R_{\nu }^{\mu }-\frac{1}{2}\,\delta _{\nu }^{\mu
}\,R+\Lambda \,\delta _{\nu }^{\mu }\,,
\end{equation}%
and the Lanczos tensor that describes contribution of the
quadratic-curvature gravitational terms,
\begin{eqnarray}
H_{\nu }^{\mu } &=&-\frac{\alpha }{8}\,\delta _{\nu \nu _{1}\cdots \nu
_{4}}^{\mu \mu _{1}\cdots \mu _{4}}\,R_{\mu _{1}\mu _{2}}^{\nu _{1}\nu
_{2}}R_{\mu _{3}\mu _{4}}^{\nu _{3}\nu _{4}}\,,  \notag \\
&=&-\frac{\alpha }{2}\,\delta _{\nu }^{\mu }\left( R^{2}-4R^{\alpha \beta
}R_{\alpha \beta }+R^{\alpha \beta \lambda \sigma }R_{\alpha \beta \lambda
\sigma }\right)  \notag \\
&&+2\alpha \left( RR_{\nu }^{\mu }-2R^{\mu \lambda }R_{\lambda \nu
}-2R_{\lambda \nu \sigma }^{\mu }R^{\lambda \sigma }+R^{\mu \alpha \lambda
\sigma }R_{\nu \alpha \lambda \sigma }\right) \,.
\end{eqnarray}%
In EGB AdS gravity, second order polynomial (\ref{effective LL}) gives two
(real) different radii square
\begin{equation}
\ell _{\text{eff}}^{(\pm )2}=\frac{2\alpha \left( d-2\right) \left(
d-3\right) }{1\pm \sqrt{1-\frac{4\alpha }{\ell ^{2}}\,\left( d-2\right)
\left( d-3\right) }}\,,  \label{Lpm_eff}
\end{equation}%
when $\alpha <\alpha _{\text{CS}}=\frac{\ell ^{2}}{4\left( d-2\right) \left(
d-3\right) }$. Then the l.h.s.~of the gravitational equations of motion (\ref%
{grav}) can be factorized,%
\begin{equation}
-\frac{\alpha }{8}\,\delta _{\nu \nu _{1}\cdots \nu _{4}}^{\mu \mu
_{1}\cdots \mu _{4}}\,\left( R_{\mu _{1}\mu _{2}}^{\nu _{1}\nu _{2}}+\frac{1%
}{\ell _{\text{eff}}^{(+)2}}\,\delta _{\mu _{1}\mu _{2}}^{\nu _{1}\nu
_{2}}\right) \left( R_{\mu _{3}\mu _{4}}^{\nu _{3}\nu _{4}}+\frac{1}{\ell _{%
\text{eff}}^{(-)2}}\,\delta _{\mu _{3}\mu _{4}}^{\nu _{3}\nu _{4}}\right)
=T_{\nu }^{\mu }\,.
\end{equation}%
Note that only the negative branch with the radius $\ell _{\text{eff}}^{(-)}$
(\textquotedblleft stringy\textquotedblright\ vacuum ) reduces to the bare
AdS radius $\ell $ when $\alpha \rightarrow 0$, whereas $\ell _{\text{eff}%
}^{(+)}$ is a new AdS vacuum typical for the EGB gravity. A linearized
theory around the stringy vacuum shows a presence of the ghosts \cite%
{Boulware:1985wk} indicating that this\ vacuum might be false, but the
unstable modes are not excited by the spherically symmetric black hole \cite%
{Deser:2002rt}. Thus, because we are looking at the
thermal CFTs dual to spherical black holes, we shall allow, in principle,
both vacua in our discussion.

On the other hand, we shall omit the Chern-Simons point, $\alpha _{\text{CS}%
} $, in the space of parameters, where the theory has the unique AdS radius (%
$\ell _{\text{eff}}^{(-)2}=\ell _{\text{eff}}^{(+)2}=\ell ^{2}/2$) and the
AdS vacuum is two-fold degenerate, because in that point the EGB
Lagrangian becomes a Chern-Simons AdS gravity \cite{Chamseddine:1990gk} that
has augmented number of gauge symmetries and has to be treated independently
\cite{Miskovic:2005di}.

To study black holes, we assume a static, maximally symmetric metric in the
local coordinates $x^{\mu }=(t,r,y^{m})$, $m=2,\ldots d,$
\begin{equation}
ds^{2}=g_{\mu \nu }(x)\,dx^{\mu }dx^{\nu }=-f(r)\,dt^{2}+\frac{dr^{2}}{%
f(r)N(r)}+r^{2}\gamma _{mn}(y)\,dy^{m}dy^{n}\,.  \label{BH metric}
\end{equation}%
The radial coordinate is chosen so that the boundary is placed at radial
infinity, $r\rightarrow \infty $, and parameterized by $x^{i}=(t,y^{m})$.
The metric $\gamma _{nm}$ of the transversal section $r=Const$ describes a $%
(d-1)$-dimensional space with the constant curvature $k=1,0$ or $-1$,
corresponding to spherical, flat or hyperbolic geometry, respectively. In
particular, $k$ determines the geometry of an event horizon of the black
hole, $r_{+}$, defined as the largest root of the equation $f(r_{+})=0$. We
are interested in the outer region only, $r\geq r_{+}$, where $f(r)\geq 0$.
The function $N(r)>0$ is finite at the horizon.

In order to have flatter\ boundary, that is suitable for a holographic
description of a fluid, we shall restrict our study to planar AdS$_{d+1}$
black holes with noncompact horizons, $k=0$, whose transversal section is $%
\mathbb{R}^{d-1}$ with the flat metric $\gamma _{mn}=\delta _{mn}$. We also
require that the curvature is slowly varying, that is, the black hole
horizon is big, $r_{+}\gg \ell _{\text{eff}}$.

Planar asymptotically AdS spacetimes have the metric functions that behave
for large $r$ as,
\begin{eqnarray}
f(r) &\rightarrow &\frac{r^{2}}{\ell _{\text{eff}}^{2}}+\mathcal{O}\left(
1/r\right) \,,  \notag \\
f(r)N(r) &\rightarrow &\frac{r^{2}}{\ell _{\text{eff}}^{2}}+\mathcal{O}%
\left( 1/r\right) \,,  \label{AAdS planar}
\end{eqnarray}%
independently on how fast $\mathcal{O}\left( 1/r\right) $ tends to zero. The
Hawking temperature of the black hole (\ref{BH metric}) in this spacetime
reads%
\begin{equation}
T=\frac{1}{4\pi }\,f^{\prime }(r_{+})\sqrt{N(r_{+})}\,,  \label{THawking}
\end{equation}%
and it increases linearly with $r_{+}$, so that the large horizon
approximation corresponds to high temperatures in a holographically dual
field theory.

We also assume that the black hole is electrically charged, with an Abelian
gauge field that has the same isometries as the metric (\ref{BH metric}),
\begin{equation}
A_{\mu }=\phi \left( r\right) \,\delta _{\mu }^{t}\,.  \label{A}
\end{equation}%
The field strength is expressed in terms of the electric field $E(r)=-\phi
^{\prime }(r)$ as $F_{\mu \nu }=E(r)\,\delta _{\mu \nu }^{tr}$, where the
prime denotes the radial derivative. Note that $F^{2}=-2NE^{2}$.

The unknown functions $f(r)$, $N(r)$, $\phi (r)$ and $\Psi (r)$ obey the
differential equations%
\begin{eqnarray}
0 &=&\mathcal{E}_{\mu \nu }:=G_{\mu \nu }+H_{\mu \nu }-T_{\mu \nu }\,,
\notag \\
0 &=&\mathcal{E}^{\mu }:=\frac{1}{\sqrt{-g}}\,\partial _{\nu }\left( 4\sqrt{%
-g}F^{\mu \nu }\dfrac{d\mathcal{L}}{dF^{2}}\right) -\mathcal{F}g^{\mu \nu
}A_{\nu }\,,  \notag \\
0 &=&\mathcal{E}:=\frac{1}{\sqrt{-g}}\,\partial _{\mu }\left( \sqrt{-g}%
\,g^{\mu \nu }\partial _{\nu }\Psi \right) -m^{2}\Psi -\dfrac{1}{2}\,\dfrac{d%
\mathcal{F}}{d\Psi }\,g^{\mu \nu }A_{\mu }A_{\nu }\,.  \label{KG}
\end{eqnarray}%
For sufficiently high temperatures, these equations possess a solution
without scalar field, $\Psi =0$, that is dual to a holographic QFT without
condensate. Its most general form for an arbitrary GB coupling $\alpha $ and
the NED Lagrangian $\mathcal{L}(F^{2})$ was found in Ref.\cite%
{Miskovic:2010ui},%
\begin{equation}
f(r)=\frac{r^{2}}{2\alpha \left( d-2\right) \left( d-3\right) }\left[ 1\pm
\sqrt{1-4\alpha \left( d-2\right) \left( d-3\right) \left( \frac{1}{\ell ^{2}%
}-\frac{\mu }{r^{d}}+\frac{2\mathcal{T}(q,r)}{\left( d-1\right) r^{d}}%
\right) }\right] \,,  \label{f_sol}
\end{equation}%
where $\mu $ is an integration constant related to the black hole mass, and
the positive function $\mathcal{T}(q,r)$ corresponds to the total matter
energy in the region between the horizon and the distance of radius $r$,
\begin{equation}
\mathcal{T}(q,r)=\frac{2}{d}\,\left. \left( \frac{1}{4}\,r^{d}\mathcal{L}%
-qrE+\left( d-1\right) q\phi \right) \right\vert _{r_{+}}^{r}.
\end{equation}%
When $\Psi =0$, the electric field is calculated from the algebraic equation
\begin{equation}
E\left. \frac{d\mathcal{L}}{dF^{2}}\right\vert _{F^{2}=-2E^{2}}=-\frac{q}{%
r^{d-1}}\,.  \label{r(E)}
\end{equation}%
In the special case of the Born-Infeld Lagrangian $\mathcal{L}(F^{2})$ \cite%
{Born:1934gh}, the solution and its thermodynamics were discussed in Ref.
\cite{Miskovic:2008ck}.

When $\Psi \neq 0$, Eqs.(\ref{KG}) become a system of nonlinear differential
equations. With help of the identities given in Appendix \ref{TBH}, we write
them in components as%
\begin{eqnarray}
\mathcal{E}_{r}^{r} &=&\dfrac{d-1}{2}\left[ \left( d-2\right) \,\frac{Nf}{%
r^{2}}+\frac{Nf^{\prime }}{r}-2\alpha \,\left( d-2\right) \left( d-3\right)
N^{2}\left( \frac{ff^{\prime }}{r^{3}}+\frac{d-4}{2}\frac{f^{2}}{r^{4}}%
\right) -\frac{d}{\ell ^{2}}\right] -T_{r}^{r}\,,  \notag \\
\mathcal{E}_{t}^{t} &=&\mathcal{E}_{r}^{r}+\dfrac{N^{\prime }f}{2r}\,\left(
d-1\right) \left( 1-2\alpha \,\left( d-2\right) \left( d-3\right) \,\frac{Nf%
}{r^{2}}\right) +\frac{1}{2}\,fN\,\Psi ^{\prime 2}+\frac{\mathcal{F}\phi ^{2}%
}{2f}\,,  \label{diff_f}
\end{eqnarray}%
where the matter energy-momentum tensor contributes through%
\begin{equation}
T_{r}^{r}=\frac{1}{2}\,\mathcal{L}+2NE^{2}\,\frac{d\mathcal{L}}{dF^{2}}+%
\frac{\mathcal{F}\phi ^{2}}{4f}-\frac{1}{4}\,m^{2}\Psi ^{2}+\frac{1}{4}%
\,fN\Psi ^{\prime }{}^{2}\,.  \label{Trr}
\end{equation}%
One can also check that, as an effect of a backreaction of the gravitational
field, the components $T_{t}^{t}$ and $T_{r}^{r}$ differ, that is,
\begin{equation}
T_{t}^{t}=T_{r}^{r}-\frac{1}{2}\,fN\,\Psi ^{\prime 2}-\frac{\mathcal{F}\phi
^{2}}{2f}\,.  \label{Ttt}
\end{equation}%
The presence of the scalar field decreases total matter energy
density (because $f$, $N$, $\mathcal{F}>0$ outside the horizon). This means
that, if a solution with the scalar field exists, then we can expect that it
would decrease the total energy of the system and, therefore, be
energetically more favorable, producing a phase transition.

The scalar field equation (\ref{KG}) in the chosen ansatz reads
\begin{equation}
\mathcal{E}=\frac{\sqrt{N}}{r^{d-1}}\,\left( r^{d-1}\sqrt{N}f\,\Psi ^{\prime
}\right) ^{\prime }-m^{2}\Psi +\dfrac{d\mathcal{F}}{d\Psi }\dfrac{\phi ^{2}}{%
2f}\,.  \label{KG_eval}
\end{equation}%
In the NED equation in (\ref{KG}), only the time-like component is
non-trivial,
\begin{equation}
\mathcal{E}^{t}=-\frac{\sqrt{N}}{r^{d-1}}\left( 4r^{d-1}\sqrt{N}E\,\dfrac{d%
\mathcal{L}}{dF^{2}}\right) ^{\prime }+\frac{\mathcal{F}\phi }{f}\,.
\label{Maxwell_t}
\end{equation}%
We define the electric potential at distance $r$ measured with respect to
the horizon $r_{+}$ as%
\begin{equation}
\phi (r)=-\int\limits_{r_{+}}^{r}ds\,E(s)\,.  \label{potential}
\end{equation}%
The quantity of physical interest is the chemical potential $\Phi =\phi
(\infty )-\phi (r_{+})$, or the potential at infinity measured with respect
to the event horizon,%
\begin{equation}
\Phi =\phi (\infty )=-\int\limits_{r_{+}}^{\infty }ds\,E(s)\,.  \label{Phi}
\end{equation}%
This choice of the reference point satisfies $\phi (r_{+})=0$, since any
other referent point would lead to nonvanishing $\phi (r_{+})$ and a
negative effective mass of the scalar field \cite{Gubser:2008px}. Indeed,
the electric potential couples to the scalar field so that, from Eq.(\ref%
{KG_eval}), it contributes to its effective mass as $m_{\text{eff}%
}^{2}=m^{2}-\frac{\phi ^{2}}{f}\left. \frac{d\mathcal{F}}{d\Psi ^{2}}%
\right\vert _{\Psi =0}$. This mass can be divergent on the horizon unless we
impose $\phi =0$ there. It is also worthwhile noticing that the effective
potential for the scalar field, leading to the equation of motion (\ref%
{KG_eval}), has the form%
\begin{equation}
V_{\text{eff}}=\frac{1}{2}\,m^{2}\Psi ^{2}-\frac{\phi ^{2}}{2f}\,\mathcal{F}%
(\Psi )\,,
\end{equation}%
and because $\mathcal{F}$ is positive, the interaction decreases the
effective potential and therefore the total energy of the system. The
systems with unbounded potentials in gravity are, in general, known to lead
to hairy black hole solutions.

Finally, it is straightforward to check that the gravitational equation $%
\mathcal{E}_{m}^{n}$ is not independent. Using the expressions (\ref{Hmm})
given in Appendix \ref{TBH}, we find%
\begin{equation}
\mathcal{E}_{n}^{m}=\delta _{n}^{m}\left[ \frac{\left( r^{d-1}\mathcal{E}%
_{r}^{r}\right) ^{\prime }}{r^{d-2}}+\frac{rf^{\prime }}{2f}\,\left(
\mathcal{E}_{r}^{r}-\mathcal{E}_{t}^{t}\right) +\frac{r}{2}\,\left( \Psi
^{\prime }\mathcal{E}-E\,\mathcal{E}^{t}\right) \right] \,,
\end{equation}%
and therefore this equation always vanishes.

The field equations (\ref{diff_f}), (\ref{KG_eval}) and (\ref{Maxwell_t})
are at most\ second order differential equations in $\left\{ f,N,\phi ,\Psi
\right\} $, defined between the horizon and the asymptotic boundary. Thus,
for each field we have to impose at most two boundary conditions, at $r_{+}$
and $r\rightarrow \infty $. Asymptotic sector we shall discuss later. As
respect to $r_{+}$, we require that all fields and their derivatives are
finite on the horizon, as well as $f(r_{+})=0$\thinspace , $\phi (r_{+})=0$
and $f^{\prime }(r_{+})$ fixed by the temperature. Then the values of other
fields and their derivatives at $r_{+}$ can be easily deduced from the
equations of motion. The scalar field, for example, has to satisfy
\begin{equation}
m^{2}\Psi (r_{+})=4\pi T\sqrt{N(r_{+})}\,\Psi ^{\prime }(r_{+})\,,
\end{equation}%
and $E\mathcal{(}r_{+}\mathcal{)}$ and $N^{\prime }(r_{+})$ are obtained
from (\ref{diff_f}) and (\ref{Maxwell_t}).

For an arbitrary $r$, the field equations are not exactly soluble when $\psi
\neq 0$. To deal with them analytically, one can use the matching method,
that was first applied to superconductors with higher-order corrections in
Ref.\cite{Gregory:2009fj}. The method consists in finding two approximative
solutions to the equations of motion in the leading order: one in the
vicinity of the horizon and another in the asymptotic region. These two
solutions are matched smoothly at the arbitrary intermediate point. This
technique allows to obtain an analytic expression for the critical
temperature. The results depend on the matching point parameter $z_{m}$,
even though its presence does not change quantitatively the features of a
phase transition. Another analytic method found in the literature is the
Sturm-Liouville one, that was used to study holographic superconductors in
Ref.\cite{Siopsis:2010uq}.

In our approach, we shall use only an asymptotic expansion. When the gravity
action is finite in the asymptotic region, its expansion in the vicinity of
the AdS boundary can be viewed as a holographic reconstruction of the bulk
fields (metric and matter fields) starting from the boundary field data
\cite{de Haro:2000xn}. This method cannot be used to fix all parameters in
the expansion because there always remain some unfixed coefficients at a
given order, depending on the dimension. This ambiguity is well-understood
in the Fefferman-Graham coordinate frame \cite{Fefferman-Graham} as coming
from non-local terms in the quantum effective action, precisely the ones
related to the holographic stress tensor \cite{Schwimmer:2000cu}. Even with
this ambiguity, the subleading orders encode the information on the
existence of phase transitions for a given set of the coupling constants.
When the transition exists, they enable to calculate the critical
temperature and the critical exponent in an analytic way.

In order to develop the above ideas in detail, first we have to remove the
infrared divergences in the action. Then we shall construct the on-shell action that will yield, in the Euclidean section, the thermodynamic
potential of the black hole, that is identified with a free energy in a
holographic QFT. This result can be obtained exactly.

\section{Euclidean bulk action}

We evaluate the bulk action (\ref{I0}) using the equations of motion and
show that it is divergent. The Euclidean action $I_{0}^{E}=-iI_{0}=\int
d^{d+1}x\sqrt{-g}\,\mathcal{L}_{0}$ is obtained by the Wick rotation of the
time coordinate, $t=-i\tau $, where the Euclidean time $\tau $ is periodic,
with the period $T^{-1}$, in order to avoid a conical singularity at the
horizon. Assuming that the bulk Lagrangian $\mathcal{L}_{0}$, evaluated
on-shell for a static and maximally symmetric solution, depends only on the
radial coordinate, we obtain%
\begin{equation}
I_{0}^{E}=-\frac{V_{d-1}}{T}\int\limits_{r_{+}}^{\infty }dr\,\frac{r^{d-1}}{%
\sqrt{N}}\,\mathcal{L}_{0}(r)\,,
\end{equation}%
where $V_{d-1}=\int d^{d-1}y\,\sqrt{\gamma }$ is the (infinite) volume of
the flat transversal section. The physical quantity is an action per unit
volume.

Using the gravitational equation of motion (\ref{diff_f}), the
EGB term becomes%
\begin{equation}
\mathcal{L}_{\text{EGB}}=-\frac{\sqrt{N}}{r^{d-1}}\left[ \rule{0pt}{15pt}%
\sqrt{N}\,f^{\prime }\left( r^{d-1}-2\alpha \left( d-1\right) \left(
d-2\right) r^{d-3}Nf\right) \right] ^{\prime }-2T_{t}^{t}\,.  \label{L_EGB}
\end{equation}%
Furthermore, the NED Lagrangian density, $\mathcal{L}$, and the scalar field
Lagrangian density, $\mathcal{L}_{\text{S}}$, can be calculated directly
from the energy momentum-tensor (\ref{Ttt}),%
\begin{equation}
\mathcal{L+L}_{\text{S}}=2T_{t}^{t}-4NE^{2}\,\frac{d\mathcal{L}}{dF^{2}}+%
\frac{\mathcal{F}\phi ^{2}}{f}\,.
\end{equation}%
Summing up all contributions, the bulk Euclidean action becomes%
\begin{eqnarray}
I_{0}^{E} &=&\frac{V_{d-1}}{2\kappa ^{2}T}\int\limits_{r_{+}}^{\infty }dr\,%
\left[ \rule{0pt}{15pt}r^{d-1}\sqrt{N}\,f^{\prime }\left( 1-2\alpha \left(
d-1\right) \left( d-2\right) \frac{Nf}{r^{2}}\right) \right] ^{\prime }
\notag \\
&&\qquad +\frac{V_{d-1}}{2\kappa ^{2}T}\int\limits_{r_{+}}^{\infty
}dr\,r^{d-1}\left( 4\sqrt{N}E^{2}\,\frac{d\mathcal{L}}{dF^{2}}-\frac{%
\mathcal{F}\phi ^{2}}{f\sqrt{N}}\right) \,,
\end{eqnarray}%
where we omit writing that $\frac{d\mathcal{L}}{dF^{2}}$ is evaluated at $%
F^{2}=-2NE^{2}$. In order to show that the second line in the above equation
is also a total derivative, we use the electromagnetic equation of motion (%
\ref{Maxwell_t}) and write%
\begin{equation}
r^{d-1}\left( 4\sqrt{N}E^{2}\,\dfrac{d\mathcal{L}}{dF^{2}}-\frac{\mathcal{F}%
\phi ^{2}}{f\sqrt{N}}\right) =-\left( 4r^{d-1}\sqrt{N}\phi E\,\dfrac{d%
\mathcal{L}}{dF^{2}}\right) ^{\prime }\,,
\end{equation}%
and we arrive to the final form of the bulk action,%
\begin{equation}
I_{0}^{E}=\frac{V_{d-1}}{2\kappa ^{2}T}\,\left. \left[ r^{d-1}\sqrt{N}%
\,f^{\prime }\left( 1-2\alpha \left( d-1\right) \left( d-2\right) \frac{Nf}{%
r^{2}}\right) -4r^{d-1}\sqrt{N}\,\phi E\,\dfrac{d\mathcal{L}}{dF^{2}}\right]
\right\vert _{r_{+}}^{\infty }\,.  \label{I0_E}
\end{equation}%
This expression is valid for any charged black hole solution, independently
on its explicit form, for all NED and scalar interactions $\mathcal{L}$ and $%
\mathcal{F}$, and any value of the GB constant $\alpha $.

Eq.(\ref{I0_E}) is clearly divergent for the global AdS space, and therefore
it is IR divergent in asymptotically AdS sector, and has to be regularized
by a suitable addition of boundary terms.

\section{Variational principle and boundary terms \label{Var}}

An action functional is well-defined if it satisfies the finite action
principle, that is, it is differentiable upon taking variational derivatives
in the fields, and free of IR divergences. This means that the action has to
reach an extremum for a given set of boundary conditions. Removal of the
divergences in the asymptotic region can be achieved if one supplements the
boundary term $B$ to the bulk action $I_{0}$, so that the total action $%
I=I_{0}+B$ fulfills the above conditions. As a consequence, the finite total action
also ensures the finiteness of the Noether charges and the Euclidean action,
that is necessary for studying the black hole thermodynamics.

The pure gravitational part of the bulk action $I_{0}$ does not fulfill the
above conditions. Indeed, the on-shell boundary terms obtained from the
variation of Eq.(\ref{I0}) can be written, using the Stokes' theorem in the
spacetime with a boundary whose an outward pointing unit normal is $n_{\mu }$%
, as
\begin{eqnarray}
\delta I_{0} &=&\frac{1}{2\kappa ^{2}}\,\int d^{d}x\,\sqrt{-h}\,n_{\mu }%
\left[ -\delta _{\nu \nu _{1}\nu _{2}\nu _{3}}^{\mu \mu _{1}\mu _{2}\mu
_{3}}\,g^{\nu \alpha }\delta \Gamma _{\mu _{1}\alpha }^{\nu _{1}}\left(
\alpha R_{\mu _{2}\mu _{3}}^{\nu _{2}\nu _{3}}+\frac{1}{\left( d-1\right)
\left( d-2\right) }\,\delta _{\mu _{2}}^{\nu _{2}}\delta _{\mu _{3}}^{\nu
_{3}}\right) \right.  \notag \\
&&\qquad +\left. \delta A_{\nu }\,4F^{\mu \nu }\frac{d\mathcal{L}}{dF^{2}}%
-\delta \Psi \nabla ^{\mu }\Psi -\mathcal{F\,}\delta p\,\left( \nabla ^{\mu
}p-A^{\mu }\right) \rule{0pt}{15pt}\right] \,.  \label{theta}
\end{eqnarray}%
In order to identify the boundary quantities in a simple way, it is
convenient to choose the local coordinates as $x^{\mu }=(r,x^{i})$, where $%
x^{i}$ ($i=2,\ldots d$) parameterize the boundary placed at $r=Const$, so
that the normal vector has the form $n_{\mu }=\left( \mathcal{N}(r),0\right)
$. This choice of $n_{\mu }$ defines a Gauss-normal frame%
\begin{equation}
ds^{2}=\mathcal{N}^{2}\left( r\right) \,dr^{2}+h_{ij}(r,x)\,dx^{i}dx^{j}\,,
\label{normal metric}
\end{equation}%
where both the lapse function $\mathcal{N}(r)$ and the induced boundary
metric $h_{ij}(r,x)$ can be related straightforwardly to the metric
functions $f(r)$ and $N(r)$ used in the black hole ansatz (\ref{BH metric}),%
\begin{equation}
\mathcal{N}^{2}=\frac{1}{fN}\,,\qquad h_{ij}=\left(
\begin{array}{cc}
-f & 0 \\
0 & r^{2}\delta _{mn}%
\end{array}%
\right) \,,\qquad \sqrt{-h}=r^{d-1}\sqrt{f}\,.  \label{Nh}
\end{equation}%
The extrinsic curvature of the boundary is defined as a covariant derivative
of the unit normal, $K_{\mu \nu }=-\nabla _{\mu }n_{\nu }$. In a Riemann
space it is symmetric, and in the ansatz (\ref{BH metric}) it has only
non-vanishing components $K_{ij}$, or%
\begin{equation}
K_{j}^{i}=h^{ik}K_{kj}=-\frac{1}{2}\,\sqrt{Nf}\,h^{ik}h_{kj}^{\prime
}=\left(
\begin{array}{cc}
-\frac{f^{\prime }}{2}\sqrt{\frac{N}{f}} & 0 \\
0 & -\frac{1}{r}\sqrt{Nf}\,\delta _{n}^{m}%
\end{array}%
\right) \,.  \label{Kbh}
\end{equation}

In the Gaussian frame, a variation of the Christoffel symbol gives rise to
the terms proportional to both the variation of the induced metric, $\delta
h_{ij}$, and the variation of the extrinsic curvature, $\delta K_{ij}$. This
action, clearly, it is not differentiable, since $\delta I_{0}/\delta h_{ij}$
is not well-defined. The differentiability of the gravitational action for
the Dirichlet boundary conditions on the induced metric can be recovered by
addition of the Gibbons-Hawking term. However, obtained Dirichlet action
still remains IR divergent. To heal these divergences, one should apply one
of known regularization methods suitable for application of the holographic
principle, i.e., that is covariant and background-independent. One
possibility is to use the holographic renormalization \cite{de
Haro:2000xn,Henningson:1998gx}, that is systematic, but technically involved
procedure in higher dimensions, where a complete counterterm series has not
been written explicitly yet.

We turn, therefore, to an alternative expression for a boundary term that
regularizes gravity action, called Kounterterms because it depends
explicitly on the extrinsic curvature $K_{ij}$ . An advantage of this
procedure is that the boundary term is known for any LL
gravity in any dimension because its form is universal, of a geometric
origin \cite{Kofinas:2007ns}. Additionally, it is background
independent, covariant and in Fefferman-Graham coordinates gives the results
consistent with the holographic renormalization \cite{Miskovic:2007mg}, in
cases when the last one can be done explicitly.

The Kounterterms have different form in even and odd dimensions. In even
dimensions $D=2n>4$, the gravitational part of the boundary term corresponds
to the $n$-th Chern form in $d=2n-1$ boundary dimensions,%
\begin{equation}
B_{\text{EGB,}2n-1}=c_{2n-1}\int d^{2n-1}x\sqrt{-h}\,\int\limits_{0}^{1}du\,%
\delta _{i_{1}\cdots i_{2n-1}}^{j_{1}\cdots
j_{2n-1}}\,K_{j_{1}}^{i_{1}}\Sigma _{j_{2}j_{3}}^{i_{2}i_{3}}(u)\cdots
\Sigma _{j_{2n-2}j_{2n-1}}^{i_{2n-2}i_{2n-1}}(u)\,,  \label{B_G}
\end{equation}%
where the constant $c_{2n-1}$ reads \cite{Kofinas:2006hr}%
\begin{equation}
c_{2n-1}=-\frac{(-\ell _{\text{eff}}^{2})^{n-1}}{\kappa ^{2}\left(
2n-2\right) !}\left( 1-\frac{2\alpha }{\ell _{\text{eff}}^{2}}\,\left(
2n-2\right) \left( 2n-3\right) \right) \,.
\end{equation}%
The tensor $\Sigma _{j_{1}j_{2}}^{i_{1}i_{2}}=\mathcal{R}%
_{j_{1}j_{2}}^{i_{1}i_{2}}-u^{2}\left(
K_{j_{1}}^{i_{1}}K_{j_{2}}^{i_{2}}-K_{j_{2}}^{i_{1}}K_{j_{1}}^{i_{2}}\right)
$ has symmetries of the Riemann tensor and it is constructed from the
intrinsic boundary curvature $\mathcal{R}_{\ jkl}^{i}(h)$ and the extrinsic
curvature $K_{j}^{i}$. With the boundary term (\ref{B_G}), the action
principle for the gravitational fields is well-posed if the spacetime is
asymptotically locally AdS,%
\begin{equation}
R_{j_{1}j_{2}}^{i_{1}i_{2}}+\frac{1}{\ell _{\text{eff}}^{2}}\,\delta
_{j_{1}j_{2}}^{i_{1}i_{2}}=0\text{\thinspace ,\qquad on }\partial \mathcal{M}%
\,.
\end{equation}%
For the flat transversal section, $\mathcal{R}_{\ jkl}^{i}$ vanishes in $B_{%
\text{EGB},2n-1}$ and the parametric integral can be solved exactly,%
\begin{equation}
B_{\text{EGB,}2n-1}=\left( -1\right) ^{n-1}\left( 2n-2\right)
!\,c_{2n-1}\,\int d^{2n-1}x\,\sqrt{-h}\det K_{j}^{i}\,.\,
\end{equation}%
Using (\ref{Nh}) and (\ref{Kbh}), the determinant of the extrinsic curvature
is evaluated as
\begin{equation}
\det K_{j}^{i}=\frac{1}{d!}\,\delta _{j_{1}\cdots j_{d}}^{i_{1}\cdots
i_{d}}\,K_{i_{1}}^{j_{1}}\cdots K_{i_{d}}^{j_{d}}=-N^{n-\frac{1}{2}}\,\frac{%
f^{\prime }f^{n-\frac{3}{2}}}{2r^{^{2n-2}}}\,.
\end{equation}%
Similarly, the Euclidean gravitational boundary term in even dimensions,
calculated in the chosen ansatz, reads%
\begin{equation}
B_{\text{EGB},2n-1}^{E}=\frac{V_{2n-2}}{2T}\,\left( -1\right) ^{n}\left(
2n-2\right) !\,c_{2n-1}\lim_{r\rightarrow \infty }\left( N^{n-\frac{1}{2}%
}f^{n-1}f^{\prime }\right) \,.  \label{B_even}
\end{equation}

In odd dimensions $D=2n+1$, a universal boundary term is derived from the
second fundamental form, and its expression given in terms of a double
parametric integration reads%
\begin{equation}
B_{\text{EGB},2n}=c_{2n}\int d^{2n}x\sqrt{-h}\int\limits_{0}^{1}du\int%
\limits_{0}^{u}ds\,\delta _{i_{1}\cdots i_{2n}}^{j_{1}\cdots
j_{2n}}\,K_{j_{1}}^{i_{1}}\delta _{j_{2}}^{i_{2}}\,\Sigma
_{j_{3}j_{4}}^{i_{3}i_{4}}(u,s)\cdots \Sigma
_{j_{2n-1}j_{2n}}^{i_{2n-1}i_{2n}}(u,s)\,.  \label{B_G2}
\end{equation}%
Now the matrix with the symmetries of the Riemann tensor is given by the
expression $\Sigma _{j_{1}j_{2}}^{i_{1}i_{2}}=\mathcal{R}%
_{j_{1}j_{2}}^{i_{1}i_{2}}-u^{2}\left(
K_{j_{1}}^{i_{1}}K_{j_{2}}^{i_{2}}-K_{j_{2}}^{i_{1}}K_{j_{1}}^{i_{2}}\right)
+\frac{s^{2}}{\ell _{\text{eff}}^{2}}\,\delta _{j_{1}j_{2}}^{i_{1}i_{2}}$,
and the coefficient $c_{2n}$ has the same as the one in Einstein-Hilbert AdS
gravity \cite{Olea:2006vd}, only the AdS radius is replaced by the effective
one,%
\begin{equation}
c_{2n}=-\frac{(-\ell _{\text{eff}}^{2})^{n-1}}{\kappa ^{2}2^{3n-3}\left(
n-1\right) !^{2}}\left( 1-\frac{2\alpha }{\ell _{\text{eff}}^{2}}\,\left(
2n-1\right) \left( 2n-2\right) \right) \,.
\end{equation}%
The action is stationary on-shell for asymptotically locally AdS spaces that
satisfy $\delta K_{ij}=0$ on $\partial \mathcal{M}$ \cite{Kofinas:2006hr}.
The last condition is equivalent to the Dirichlet boundary condition on the
induced metric, as $K_{ij}$ and $h_{ij}$ are proportional in the leading
order near the boundary.

Evaluated on the equations of motion, the Euclidean boundary term reads%
\begin{eqnarray}
B_{\text{EGB},2n}^{E} &=&-\frac{V_{2n-1}}{Tn}\,2^{n-2}\left( 2n-1\right)
!\,c_{2n}\lim_{r\rightarrow \infty }\,\sqrt{N}\left[ \left( f-\frac{%
rf^{\prime }}{2}\right) \left( -Nf+\frac{r^{2}}{\ell _{\text{eff}}^{2}}%
\right) ^{n-1}\right.  \notag \\
&&\qquad \qquad +\left. nrf^{\prime }\left( -Nf\right) ^{n-1}\left.
_{2}F_{1}\right. \left( 1-n,\frac{1}{2};\frac{3}{2};\frac{r^{2}}{\ell _{%
\text{eff}}^{2}Nf}\right) \right] \,,  \label{B_odd}
\end{eqnarray}%
where the ordinary hypergeometric function $\left. _{2}F_{1}\right. \left(
1-n,\frac{1}{2};\frac{3}{2};\frac{r^{2}}{\ell _{\text{eff}}^{2}Nf}\right) $,
represented by the hypergeometric series (see Appendix \ref{Hyper}), is just
a compact way to write the counterterm series using the expansion%
\begin{equation}
\left. _{2}F_{1}\right. \left( 1-n,\frac{1}{2};\frac{3}{2};\frac{r^{2}}{\ell
_{\text{eff}}^{2}Nf}\right) =1-\frac{n-1}{3}\,\frac{r^{2}}{\ell _{\text{eff}%
}^{2}Nf}+\frac{\left( n-1\right) \left( n-2\right) }{10}\,\frac{r^{4}}{\ell
_{\text{eff}}^{4}N^{2}f^{2}}+\cdots \,.
\end{equation}%
The above series converges when $\frac{r^{2}}{\ell _{\text{eff}}^{2}Nf}<1$.

Now we consider the Abelian gauge field and ask the action to be stationary
under its variation. The NED surface term in (\ref{theta}) vanishes when the
gauge field is held fixed on the boundary, $\delta A_{i}=0$. If the boundary
condition is chosen in that way, the Euclidean action corresponds to an
ensemble with fixed electric potential, $\phi $, and it is proportional to
the Gibbs thermodynamic potential $G(T,\phi ,\ldots )$ in grand canonical
ensemble.

We shall, however, choose the boundary term of the form%
\begin{equation}
B_{\text{NED}}=\frac{2}{\kappa ^{2}}\,\int d^{d+1}x\,\partial _{\mu }\left(
\sqrt{-g}\,A_{\nu }F^{\nu \mu }\frac{d\mathcal{L}}{dF^{2}}\right) \,.
\label{Bned}
\end{equation}%
In this case, the on-shell variation of the total NED action becomes%
\begin{equation}
\delta \left( I_{\text{NED}}+B_{\text{NED}}\right) =-\frac{2}{\kappa ^{2}}%
\,\int d^{d}x\,A_{\mu }\,\delta \left( \sqrt{-h}\,n_{\nu }F^{\nu \mu }\frac{d%
\mathcal{L}}{dF^{2}}\right) \,.  \label{varI_Ned}
\end{equation}%
When $h_{ij}$ is fixed on the boundary (the Dirichlet boundary conditions
for the metric), the gauge field $A_{\mu }$ has to satisfy the Neumann-like
boundary condition $\delta \left( F_{ri}\,\frac{d\mathcal{L}}{dF^{2}}\right)
=0$ in order to give rise a stationary action. We will show in the next
section that the electric charge $Q$ is proportional to the quantity$\sqrt{-h%
}\,n_{\nu }\,F^{\nu \mu }\frac{d\mathcal{L}}{dF^{2}}$ and, thus, keeping $%
F_{ri}\,\frac{d\mathcal{L}}{dF^{2}}$ fixed on $\partial \mathcal{M}$
describes a thermodynamic system whose electric charge is kept constant, $%
\delta Q=0$. In that case, the Euclidean action is related to the Helmholtz
free energy $F(T,Q,\ldots )$ in the canonical ensemble. A choice of
canonical or grand canonical ensemble is, therefore, governed by a choice of
the boundary terms, because its addition to the Euclidean action is seen as
the Legendre transformation $G=F-Q\phi $ of the corresponding thermodynamic
potentials. A discussion on a choice of an electromagnetic boundary term in
4D Maxwell theory coupled to EH gravity and its application to black hole
thermodynamics has been discussed in Ref. \cite{Caldarelli:1999xj}.

The on-shell evaluation of the Euclidean action of the boundary term (\ref%
{Bned}) has the form%
\begin{equation}
B_{\text{NED}}^{E}=\frac{2V_{d-1}}{\kappa ^{2}T}\,\lim_{r\rightarrow \infty
}\left( r^{d-1}\phi \sqrt{N}E\,\frac{d\mathcal{L}}{dF^{2}}\right) \,.
\label{B_NED}
\end{equation}

Finally, for the scalar field, the action is stationary when $\Psi $ and $p$
are held fixed on the boundary. We also note that the Euclidean action (\ref%
{I0_E}) does not have terms associated to scalar field, that means that
there are no divergences associated to it either, and we can safely choose $%
B_{\text{S}}=0$. Thus, we choose Dirichlet boundary conditions for the
scalar field. Note that vanishing of the scalar IR divergences is an effect
of gravitational backreaction. Namely, in the matter fields probe limit,
when this backreaction is not taken into account, the scalar field becomes
divergent and it requires additional counterterms (see, for example, Ref.%
\cite{Franco:2009yz}). Our result that the backreaction heals the
divergences is a particular feature of the chosen matter couplings -- one
should not expect the same to happen for, e.g., non-minimal interaction
between $\Psi $ and the EM field, such as $\sigma (\Psi )\mathcal{L}(F^{2})$.

In the next section, we use the Noether theorem to calculate conserved
quantities associated to the local symmetries in the theory.

\section{Conserved quantities \label{Con}}

The action is invariant under Abelian gauge transformations that act
nontrivially on the following fields,%
\begin{equation}
\delta _{\lambda }A_{\mu }=\partial _{\mu }\lambda \,,\qquad \delta
_{\lambda }p=\lambda \,,
\end{equation}%
and whose associated Noether current reads%
\begin{equation}
J^{\mu }(\lambda )=\partial _{\nu }\left( \frac{\sqrt{-g}}{2\kappa ^{2}}%
\,\lambda \,4F^{\mu \nu }\frac{d\mathcal{L}}{dF^{2}}\right) \,.
\end{equation}%
The charge density in the radial foliation is described by the component $%
J^{r}(\lambda )=\partial _{i}\left( \frac{\sqrt{-g}}{2\kappa ^{2}}\,\lambda
\,4F^{ri}\frac{d\mathcal{L}}{dF^{2}}\right) $. Applying the Stokes' theorem
to the boundary manifold at $r\rightarrow \infty $ with the metric $h_{ij}$
whose time-slice boundary $t=Const$, denoted by $\Sigma _{\infty }$, is
defined by an outward-pointing time-like unit normal $u_{i}=(u_{t},u_{m})=%
\left( -\sqrt{f},0\right) $, we obtain a general formula for the total
electric charge of the black hole,%
\begin{equation}
Q=-\frac{2V_{d-1}}{\kappa ^{2}}\,\lim_{r\rightarrow \infty }\left( r^{d-1}%
\sqrt{N}E\,\frac{d\mathcal{L}}{dF^{2}}\right) \,.  \label{Q_electric}
\end{equation}%
This form of the charge justifies the interpretation of the boundary
condition $\delta \left( F_{ri}\,\frac{d\mathcal{L}}{dF^{2}}\right) =0$
mentioned in the previous section as having $Q$ fixed.

Energy-momentum of the black hole is associated to the time-like
diffeomorphisms $\xi =\xi ^{\mu }\partial _{\mu }$ that act on the fields as
Lie derivatives,%
\begin{equation}
\begin{array}{ll}
\pounds _{\xi }g_{\mu \nu }=\nabla _{\mu }\xi _{\nu }+\nabla _{\nu }\xi
_{\mu }\,,\medskip & \pounds _{\xi }\Psi =\xi ^{\nu }\partial _{\nu }\Psi \,,
\\
\pounds _{\xi }A_{\mu }=\partial _{\mu }\left( \xi ^{\nu }A_{\nu }\right)
-\xi ^{\nu }F_{\mu \nu }\,,\qquad & \pounds _{\xi }p=\xi ^{\nu }\partial
_{\nu }p\,.%
\end{array}%
\end{equation}%
The gravitational contribution to the conserved charges in
EGB gravity coupled to NED, where the same regularization
method is used, is calculated in Ref. \cite{Miskovic:2010ui} as%
\begin{equation}
Q[\xi ]=\int\limits_{\Sigma _{\infty }}d^{d-1}y\,\sqrt{\det \left(
g_{mn}\right) }\,u_{j}\,\xi ^{i}\,q_{i}^{j}\,,
\end{equation}%
where $\xi $ is an asymptotic Killing vector. The charge density tensor $%
q_{i}^{j}(r)$ has different form in even and odd dimensions, depending on
the boundary terms. In our case, $\xi =\partial _{t}$ and $%
g_{mn}=r^{2}\delta _{mn}$, so that the total energy is%
\begin{equation}
Q[\partial _{t}]=-V_{d-1}\,\lim_{r\rightarrow \infty }r^{d-1}\sqrt{f}%
\,q_{t}^{t}\,.  \label{Q[d_t]}
\end{equation}

In even dimensions ($D=2n$), the charge density tensor is given by the
formula \cite{Miskovic:2010ui}
\begin{eqnarray}
\left( q_{2n}\right) _{i}^{j} &=&\frac{1}{2\kappa ^{2}\left( 2n-2\right)
!2^{n-2}}\,\delta _{i_{1}i_{2}\cdots i_{2n-1}}^{j\,j_{2}\cdots
j_{2n-1}}\,K_{i}^{i_{1}}\times  \notag \\
&&\times \left[ \rule{0in}{0.24in}\left( \delta
_{j_{2}j_{3}}^{i_{2}i_{3}}+2\alpha \left( 2n-2\right) \left( 2n-3\right)
R_{j_{2}j_{3}}^{i_{2}i_{3}}\right) \delta _{j_{4}j_{5}}^{i_{4}i_{5}}\cdots
\delta _{j_{2n-2}j_{2n-1}}^{i_{2n-2}i_{2n-1}}\right.  \notag \\
&&\left. \rule{0in}{0.24in}+\kappa ^{2}\left( 2n-2\right)
!\,c_{2n-1}\,R_{j_{2}j_{3}}^{i_{2}i_{3}}\cdots
R_{j_{2n-2}j_{2n-1}}^{i_{2n-2}i_{2n-1}}\right] \,,
\end{eqnarray}%
where the Gauss-Codazzi relation $R_{kl}^{ij}=\mathcal{R}%
_{kl}^{ij}-K_{k}^{i}K_{l}^{j}+K_{l}^{i}K_{k}^{j}$ has to be used in order to
express $q_{i}^{j}$ completely in terms of the boundary quantities.
Evaluating the above tensor in the ansatz (\ref{Nh}) and (\ref{Kbh}), the
total energy becomes
\begin{eqnarray}
Q_{2n}[\partial _{t}] &=&\frac{V_{2n-2}}{2\kappa ^{2}}\lim_{r\rightarrow
\infty }\,r^{2n-2}\sqrt{N}f^{\prime }\times  \notag \\
&&\left[ 1-2\alpha \,\left( 2n-2\right) \left( 2n-3\right) \frac{Nf}{r^{2}}%
+\kappa ^{2}\left( 2n-2\right) !\,c_{2n-1}\left( -\frac{Nf}{r^{2}}\right)
^{n-1}\right] \,.
\end{eqnarray}

On the other hand, in odd dimensions ($D=2n+1$), the charge density tensor
contains the following terms \cite{Miskovic:2010ui},
\begin{eqnarray}
q_{i}^{j} &=&(q_{\text{vac}})_{i}^{j}+\frac{1}{2\kappa ^{2}\left(
2n-1\right) !2^{n-2}}\,\delta _{i_{1}\cdots i_{2n}}^{jj_{2}\cdots
j_{2n}}\,K_{i}^{i_{1}}\delta _{j_{2}}^{i_{2}}\times  \notag \\
&&\times \left[ \rule{0in}{0.26in}\left( \delta
_{j_{3}j_{4}}^{i_{3}i_{4}}+2\alpha \left( 2n-1\right) \left( 2n-2\right)
\,R_{j_{3}j_{4}}^{i_{3}i_{4}}\right) \delta _{j_{5}j_{6}}^{i_{5}i_{6}}\cdots
\delta _{j_{2n-1}j_{2n}}^{i_{2n-1}i_{2n}}\right.  \notag \\
&&+\left. 2\kappa ^{2}\left( 2n-1\right)
!\,nc_{2n}\int\limits_{0}^{1}du\left( R_{j_{3}j_{4}}^{i_{3}i_{4}}+\frac{u^{2}%
}{\ell _{\text{eff}}^{2}}\,\delta _{j_{3}j_{4}}^{i_{3}i_{4}}\right) \cdots
\left( R_{j_{2n-1}j_{2n}}^{i_{2n-1}i_{2n}}+\frac{u^{2}}{\ell _{\text{eff}%
}^{2}}\,\delta _{j_{2n-1}j_{2n}}^{i_{2n-1}i_{2n}}\right) \right] \,,
\end{eqnarray}%
where%
\begin{eqnarray}
(q_{\text{vac}})_{i}^{j} &=&2^{n-2}c_{2n}\,\int\limits_{0}^{1}du\,u\,\delta
_{ki_{2}\cdots i_{2n}}^{jj_{2}\cdots j_{2n}}\left( K_{i}^{k}\delta
_{j_{2}}^{i_{2}}+K_{j_{2}}^{k}\delta _{i}^{i_{2}}\right) \left( \frac{1}{2}\,%
\mathcal{R}%
_{j_{3}j_{4}}^{i_{3}i_{4}}-u^{2}K_{j_{3}}^{i_{3}}K_{j_{4}}^{i_{4}}+\frac{%
u^{2}}{\ell _{\text{eff}}^{2}}\,\delta _{j_{3}}^{i_{3}}\delta
_{j_{4}}^{i_{4}}\right) \times \cdots  \notag \\
&&\qquad \qquad \cdots \times \left( \frac{1}{2}\,\mathcal{R}%
_{j_{2n-1}j_{2n}}^{i_{2n-1}i_{2n}}-u^{2}K_{j_{2n-1}}^{i_{2n-1}}K_{j_{2n}}^{i_{2n}}+%
\frac{u^{2}}{\ell _{\text{eff}}^{2}}\,\delta _{j_{2n-1}}^{i_{2n-1}}\delta
_{j_{2n}}^{i_{2n}}\right) \,,
\end{eqnarray}%
and again we have to use the Gauss-Codazzi relation for $R_{kl}^{ij}$ to
express the tensor in terms of the boundary quantities. Note that the charge
density does not vanish identically for global AdS, $%
R_{j_{2}j_{3}}^{i_{2}i_{3}}=-\frac{1}{\ell _{\text{eff}}^{2}}\,\delta
_{j_{2}j_{3}}^{i_{2}i_{3}}$, because the contribution $(q_{\text{vac}%
})_{i}^{j}$ accounts for the vacuum energy of AdS space. The vacuum energy
depends only on the properties of the \textquotedblleft
empty\textquotedblright\ space, that is, the topological parameter $k$, the
effective AdS radius and the gravitational couplings $\kappa ^{2}$,$\alpha $%
, thus the presence of the scalar field does not influence on it. The
formula for the vacuum energy of GB black holes in asymptotically AdS space
was evaluated in Ref.\cite{Kofinas:2006hr}, but because in our case the
topological parameter is zero ($k=0$), it vanishes. This can be seen
explicitly. Evaluating $q_{t}^{t}$ in our ansatz, we obtain%
\begin{equation}
q_{t}^{t}=(q_{\text{vac}})_{t}^{t}+\frac{K_{t}^{t}}{\kappa ^{2}}\,\left[
1-2\alpha \left( 2n-1\right) \left( 2n-2\right) \,\frac{Nf}{r^{2}}+2\kappa
^{2}\left( 2n-1\right) !\,2^{n-2}c_{2n}\int\limits_{0}^{1}du\left( -\frac{Nf%
}{r^{2}}+\frac{u^{2}}{\ell _{\text{eff}}^{2}}\right) ^{n-1}\right] \,,
\end{equation}%
where the integral can be expressed in terms of the hypergeometric function,
as explained in Appendix \ref{Hyper}. Now we can show that the vacuum energy,%
\begin{equation}
(q_{\text{vac}})_{t}^{t}=\frac{2^{n-2}}{n}\,\left( 2n-1\right) !\,c_{2n}%
\sqrt{\frac{N}{f}}\left( \frac{f}{r}-\frac{f^{\prime }}{2}\right) \left( -%
\frac{Nf}{r^{2}}+\frac{1}{\ell _{\text{eff}}^{2}}\right) ^{n-1},
\end{equation}%
does not contribute to the total energy $Q[\partial _{t}]$. For planar black
holes in asymptotically AdS spacetime, the metric functions $f$ and $N$
satisfy the conditions (\ref{AAdS planar}). Evaluating the limit of $(q_{%
\text{vac}})_{t}^{t}$ in (\ref{Q[d_t]}), we get
\begin{equation}
E_{\text{vac}}=-V_{2n-1}\,\frac{2^{n-2}}{n}\,\left( 2n-1\right)
!\,c_{2n}\lim_{r\rightarrow \infty }\sqrt{N}\left( f-\frac{rf^{\prime }}{2}%
\right) \left( -Nf+\frac{r^{2}}{\ell _{\text{eff}}^{2}}\right) ^{n-1}=0\,,
\label{vacuum}
\end{equation}%
as expected. Thus, what remains is%
\begin{eqnarray}
Q_{2n+1}[\partial _{t}] &=&-\frac{V_{2n-1}}{\kappa ^{2}}\,\lim_{r\rightarrow
\infty }r^{2n-1}\sqrt{f}\,K_{t}^{t}\,\left[ \rule{0pt}{16pt}1-2\alpha \left(
2n-1\right) \left( 2n-2\right) \frac{Nf}{r^{2}}\right.  \notag \\
&&+\left. 2\kappa ^{2}\left( 2n-1\right) !\,2^{n-2}c_{2n}\,\left( -\frac{Nf}{%
r^{2}}\right) ^{n-1}\left. _{2}F_{1}\right. \left( 1-n,\frac{1}{2};\frac{3}{2%
};\frac{r^{2}}{\ell _{\text{eff}}^{2}Nf}\right) \right] .
\end{eqnarray}

In both even and odd dimensions $D$, total energy corresponds to the black
hole mass that can be written as%
\begin{equation}
M=\frac{V_{D-2}}{2\kappa ^{2}}\lim_{r\rightarrow \infty }\mathcal{M}%
_{D}(r)\,,  \label{M}
\end{equation}%
with the function of radial coordinate given in even dimensions by%
\begin{equation}
\mathcal{M}_{2n}(r)=r^{2n-2}\sqrt{N}f^{\prime }\left[ 1-2\alpha \,\left(
2n-2\right) \left( 2n-3\right) \frac{Nf}{r^{2}}+\kappa ^{2}\left(
2n-2\right) !\,c_{2n-1}\left( -\frac{Nf}{r^{2}}\right) ^{n-1}\right] \,,
\label{Meven}
\end{equation}%
and in odd dimensions as%
\begin{eqnarray}
\mathcal{M}_{2n+1}(r) &=&\sqrt{N}f^{\prime }\,\left[ \rule{0in}{0.28in}%
r^{2n-1}\left( 1-2\alpha \left( 2n-1\right) \left( 2n-2\right) \frac{Nf}{%
r^{2}}\right) \right.  \notag \\
&&+\left. 2\kappa ^{2}\left( 2n-1\right) !\,2^{n-2}c_{2n}\,r\left(
-Nf\right) ^{n-1}\left. _{2}F_{1}\right. \left( 1-n,\frac{1}{2};\frac{3}{2};%
\frac{r^{2}}{\ell _{\text{eff}}^{2}Nf}\right) \right] .  \label{Modd}
\end{eqnarray}%
Without scalar field, $N=1$ and the above formulas match the ones of Ref.
\cite{Miskovic:2010ui}.

It is straightforward to check that scalar and electromagnetic fields do not
contribute to the mass. We shall show explicitly that, in
the Euclidean section, addition of the boundary term $B_{\text{NED}}^{E}$ performs a Legendre
transformation of the thermodynamic potential. On the other hand, the scalar
field does not contribute to the mass because the
time-like isometry leaves the static scalar fields invariant, $\pounds %
_{\partial _{t}}\Psi =0$ and $\pounds _{\partial _{t}}p=0$, leading to the
given result. Furthermore, we do not need the counterterms for $\Psi $.
Thus, the above formula is the final, exact expression for the total energy of
black holes with hair in nonlinear GB AdS gravity interacting with the St%
\"{u}ckelberg scalar and NED field.

Now we proceed to evaluate the free energy for a chosen class of solutions.

\section{Quantum statistical relation for a GB superconductor \label{QSR_GB}}

Using the nomenclature of Ref.\cite{Gibbons:2004ai}, quantum statistical
relation is an expression for a thermodynamic potential of the system held
at fixed temperature obtained using the quantum statistical mechanics, that
is, directly from the partition function defined as exponent of the
Euclidean action. The charges within it are the Noether ones, a consequence
of the symmetries of the action, and they are expected to match the
thermodynamic charges that appear in the first law of thermodynamics. This
formula differs from the Smarr-like relation stemming from integration of
the first law.

In order to find the statistical relation, we have to calculate the
Euclidean action. Adding the contributions of the boundary terms (\ref%
{B_even}) and (\ref{B_NED}) to the bulk action (\ref{I0_E}), the
electromagnetic part cancels out at infinity and the total Euclidean action
in even dimensions reads%
\begin{eqnarray}
I^{E} &=&\frac{V_{2n-2}}{2\kappa ^{2}T}\,\lim_{r\rightarrow \infty }\mathcal{%
M}_{2n}(r)  \notag \\
&&-\frac{V_{d-1}}{2\kappa ^{2}T}\,\left. \left[ \sqrt{N}\,f^{\prime }\left(
r^{d-1}-2\alpha \left( d-1\right) \left( d-2\right) r^{d-3}Nf\right)
-4r^{d-1}\sqrt{N}\,\phi E\,\dfrac{d\mathcal{L}}{dF^{2}}\right] \right\vert
_{r=r_{+}},
\end{eqnarray}%
where we recognized the radial function $\mathcal{M}_{2n}(r)$ from Eq.(\ref%
{Meven}). The terms in the second line do not depend on a parity of the
dimension $d$. On the horizon, the functions $f$ and\ $\phi $ vanish and $N$
and $E\dfrac{d\mathcal{L}}{dF^{2}}$ are finite, so using the definition of
the Hawking temperature (\ref{THawking}) and the black hole mass (\ref{M}),
we can write the Euclidean action as%
\begin{equation}
I^{E}=T^{-1}M-S\,,  \label{Ie}
\end{equation}%
where the entropy $S=-\frac{V_{d-1}}{T}\left. \left[ r^{d-1}\sqrt{f}\,(q-q_{%
\text{vac}})_{t}^{t}\right] \right\vert _{r_{+}}$ is, as usual, the Noether
charge on the horizon, giving
\begin{equation}
S=\frac{2\pi V_{d-1}r_{+}^{d-1}}{\kappa ^{2}}\,.
\end{equation}%
Note that the GB term does not contribute to the entropy when the black hole
is planar. Additionally, the St\"{u}ckelberg scalar does not modify the area
law, as it happens for a conformally coupled scalar field $f(\Psi )R$\ or
other non-minimal interactions between $\Psi $ and curvature invariants \cite%
{Ashtekar:2003jh}.

A similar expression for the Euclidean action is found in odd dimensions, as
well. Summing up (\ref{I0_E}), (\ref{B_odd}) and (\ref{B_NED}) and
recognizing the expression for the mass (\ref{Modd}) and vacuum energy (\ref%
{vacuum}), we obtain the total black hole energy instead of the mass only,
\begin{equation}
I^{E}=T^{-1}\left( M+E_{\text{vac}}\right) -S\,,
\end{equation}%
that finally reduces to (\ref{Ie}) because $E_{\text{vac}}=0$. The quantum
statistical relation in canonical ensemble implies that the Helmholtz free
energy, $F=TI^{E}$, has the correct form of the Legendre transformation of
the internal energy,%
\begin{equation}
F=M-TS=\frac{V_{d-1}}{2\kappa ^{2}}\lim_{r\rightarrow \infty }\,\mathcal{M}%
_{D}(r)-\frac{2\pi TV_{d-1}r_{+}^{d-1}}{\kappa ^{2}}\,.  \label{free E}
\end{equation}%
Thus, we obtain the exact expression for the thermodynamic potential of a
family of black holes in EGB gravities that interact non-minimally with
matter. The functions $\mathcal{L}$ and $\mathcal{F}$ do not enter this
formula explicitly, but through the coupling constants.

\section{Free energy of a Lovelock superconductor in canonical and grand
canonical ensemble}

The result for the free energy (\ref{free E}) can be generalized to any
hairy LL AdS black hole coupled to NED and St\"{u}ckelberg field. As shown
in Ref.\cite{Kofinas:2007ns}, the LL AdS action is regularized by the same
boundary term as the one in Einstein-Hilbert or EGB gravity with negative
cosmological constant in even or odd dimensions, that is known to be
universal for any Lovelock gravity. The only quantity that changes is the
coefficient $c_{d}$, that depends on $\ell _{\text{eff}}$\ as one of the
roots of the polynomial (\ref{effective LL}). Proceeding the same as in EGB
case, the Lovelock action evaluated on the black hole ansatz has the form%
\begin{eqnarray}
I_{\text{LL}} &=&\frac{1}{2\kappa ^{2}}\int d^{D}x\sum_{p=0}^{[d/2]}\alpha
_{p}\,\frac{r^{d-1}}{\sqrt{N}}\,\frac{\left( d-1\right) !}{\left(
d-2p+1\right) !}\,\frac{N}{r^{2}}\left( -\dfrac{fN}{r^{2}}\right) ^{p-2}
\notag \\
&&\times \left[ \rule{0pt}{17pt}\right. \frac{pN^{\prime }}{2}\left( \left(
2p-1\right) ff^{\prime }+2\left( d-2p+1\right) \,\dfrac{f^{2}}{r}\right)
\notag \\
&&+\left. N\left( pff^{\prime \prime }+p\left( p-1\right) \,f^{\prime
2}+2p\left( d-2p+1\right) \,\dfrac{ff^{\prime }}{r}+\left( d-2p+1\right)
\left( d-2p\right) \,\dfrac{f^{2}}{r^{2}}\right) \rule{0pt}{17pt}\right] \,.
\end{eqnarray}%
Compared to the dynamics of the EGB gravity, when the higher-order curvature
corrections are added to the action, only the gravitational equations of
motion change,

\begin{eqnarray}
\mathcal{E}_{r}^{r} &=&\sum_{p=0}^{[d/2]}\alpha _{p}\,\frac{\left(
d-1\right) !}{2\left( d-2p\right) !}\,\left( -\dfrac{fN}{r^{2}}\right)
^{p-1}\left( p\,\dfrac{f^{\prime }N}{r}+\left( d-2p\right) \,\dfrac{fN}{r^{2}%
}\right) -T_{r}^{r}  \notag \\
\mathcal{E}_{t}^{t} &=&\mathcal{E}_{r}^{r}+\sum_{p=1}^{[d/2]}\alpha _{p}\,%
\frac{p\left( d-1\right) !}{2\left( d-2p\right) !}\,\left( -\dfrac{fN}{r^{2}}%
\right) ^{p-1}\dfrac{fN^{\prime }}{r}+\frac{1}{2}\,fN\,\Psi ^{\prime 2}+%
\frac{\mathcal{F}\phi ^{2}}{2f}\,,
\end{eqnarray}%
and they reduce to (\ref{diff_f}) in EGB case. Using these equations, the total bulk action coupled to the matter fields generalizes
Eq.(\ref{I0_E}) to
\begin{equation}
I_{0}^{E}=\frac{V_{d-1}}{2\kappa ^{2}T}\,\left. \left[ r^{d-1}\sqrt{N%
}\,f^{\prime }\sum_{p=1}^{[d/2]}\,\alpha _{p}\,\frac{p\left( d-1\right) !}{%
\left( d-2p+1\right) !}\,\left( -\dfrac{fN}{r^{2}}\right) ^{p-1}-4r^{d-1}%
\sqrt{N}\,\phi E\,\dfrac{d\mathcal{L}}{dF^{2}}\right] \right\vert
_{r_{+}}^{\infty }\,.
\end{equation}%
Now we follow the steps of Sections \ref{Var}-\ref{QSR_GB} in order to show
that the mass has the form (\ref{M}).

In even dimensions, the boundary term is given by the formula (\ref{B_G}),
where the Lovelock parameters $\alpha _{p}$ are introduced as%
\begin{eqnarray}
&&\left( \delta _{j_{2}j_{3}}^{i_{2}i_{3}}+2\alpha \left( 2n-2\right) \left(
2n-3\right) R_{j_{2}j_{3}}^{i_{2}i_{3}}\right) \delta
_{j_{4}j_{5}}^{i_{4}i_{5}}\cdots \delta
_{j_{2n-2}j_{2n-1}}^{i_{2n-2}i_{2n-1}}+\cdots  \notag \\
&=&\sum_{p=1}^{n}\alpha _{p}\,\frac{p\left( 2n-2\right) !}{(2n-2p)!}%
\,R_{j_{2}j_{3}}^{i_{2}i_{3}}\cdots
R_{j_{2p-2}j_{2p-1}}^{i_{2p-2}i_{2p-1}}\,\delta
_{j_{2p}j_{2p+1}}^{i_{2p}i_{2p+1}}\cdots \delta
_{j_{2n-1}j_{2n-1}}^{i_{2n-2}i_{2n-1}}\,,
\end{eqnarray}%
and the effective AdS radius in $c_{2n-1}$ is the one for the LL theory.
After a straightforward calculation, one arrives to a radial function that
defines the black hole mass in even dimensions,
\begin{equation}
\mathcal{M}_{2n}(r)=r^{2n-2}\sqrt{N}f^{\prime }\left[ \sum_{p=1}^{n-1}\alpha
_{p}\,\frac{p(2n-2)!}{(2n-2p)!}\,\left( -\frac{Nf}{r^{2}}\right)
^{p-1}+\kappa ^{2}\left( 2n-2\right) !\,c_{2n-1}\left( -\frac{Nf}{r^{2}}%
\right) ^{n-1}\right] \,.  \label{M_LL2n}
\end{equation}

In odd dimensions, the boundary term has the form (\ref{B_G2}) with the
following generalization,%
\begin{eqnarray}
&&\left( \delta _{j_{3}j_{4}}^{i_{3}i_{4}}+2\alpha \left( 2n-1\right) \left(
2n-2\right) \,R_{j_{3}j_{4}}^{i_{3}i_{4}}\right) \delta
_{j_{5}j_{6}}^{i_{5}i_{6}}\cdots \delta
_{j_{2n-1}j_{2n}}^{i_{2n-1}i_{2n}}+\cdots  \notag \\
&=&\sum_{p=1}^{n}\alpha _{p}\frac{p\left( 2n-1\right) !}{(2n-2p+1)!}%
\,R_{j_{3}j_{4}}^{i_{3}i_{4}}\cdots
R_{j_{2p-1}j_{2p}}^{i_{2p-1}i_{2p}}\,\delta
_{j_{2p+1}j_{2p+2}}^{i_{2p+1}i_{2p+2}}\cdots \delta
_{j_{2n-1}j_{2n}}^{i_{2n-1}i_{2n}}\,,
\end{eqnarray}%
and using the AdS radius that depend on all LL coupling constants. The
vacuum energy vanishes for the planar solutions, $E_{\text{vac}}=0$, and the
radial function that determines the black hole mass is
\begin{eqnarray}
\mathcal{M}_{2n+1}(r) &=&\sqrt{N}f^{\prime }\,\left[ \rule{0in}{0.28in}%
r^{2n-1}\sum_{p=1}^{n}\alpha _{p}\,\frac{p\left( 2n-1\right) !}{(2n-2p+1)!}%
\,\left( -\frac{Nf}{r^{2}}\right) ^{p-1}\right.  \notag \\
&&+\left. \kappa ^{2}\left( 2n-1\right) !\,2^{n-1}c_{2n}\,r\left( -Nf\right)
^{n-1}\left. _{2}F_{1}\right. \left( 1-n,\frac{1}{2};\frac{3}{2};\frac{r^{2}%
}{\ell _{\text{eff}}^{2}Nf}\right) \right] .  \label{M_LL2n+1}
\end{eqnarray}%
The Helmholtz free energy obtained from the quantum statistical relation has
the usual form,
\begin{equation}
F=\frac{V_{d-1}}{2\kappa ^{2}}\lim_{r\rightarrow \infty }\,\mathcal{M}%
_{D}(r)-\frac{2\pi TV_{d-1}r_{+}^{d-1}}{\kappa ^{2}}\,.  \label{F_LL}
\end{equation}%

At this point, it is straightforward to write an analogous result in the
grand canonical ensemble. Let us recall once more that the previous results,
calculated in the canonical ensemble, are obtained from the action that
fulfills the boundary conditions $\delta I_{\text{can}}=0$ when $T$, $Q$ and
$\Psi $ are held fixed on the boundary, and the on-shell Euclidean
action becomes $I_{\text{can}}^{E}=T^{-1}F=T^{-1}M-S$.

On the other hand, in grand canonical ensemble, the action is stationary for
the boundary conditions $\delta I_{\text{grand can}}=0$ when $T$, $\phi $
and $\Psi $ are held fixed on the boundary what, in practice, is realized by
not adding the NED boundary term (\ref{Bned}) to the bulk action. As
explained in Section \ref{Var}, its Euclidean continuation is related to the Gibbs potential
$G$ as%
\begin{equation}
I_{\text{grand can}}^{E}=T^{-1}G=T^{-1}M-S-T^{-1}Q\,\Phi \,,
\end{equation}%
that is exactly a Legendre transformation of the Helmholtz free energy. Here
the mass has the same form as before and the conjugated variables $\Phi$ and $Q $
are (see Eqs.(\ref{Phi}) and (\ref{Q_electric})),
\begin{equation}
\Phi
=\lim_{r\rightarrow \infty }\phi (r)\,,\qquad
Q=-\frac{2V_{d-1}}{\kappa ^{2}}\,\lim_{r\rightarrow \infty }\left( r^{d-1}%
\sqrt{N}E\,\frac{d\mathcal{L}}{dF^{2}}\right)\,.
\end{equation}%
Without scalar field, the above results coincide with the ones found in Refs.%
\cite{Miskovic:2010ey,Miskovic:2012zz} for the EGB case.

It is worthwhile mentioning that all presented results can be generalized to
topological black holes with non flat transversal section ($k\neq 0$), in
case one is interested to study the effects of horizon topology to the properties
of superconductors \cite{Myung:2010rb}.

\section{Discussion: from the free energy to the phase transition}

The Helmholz free energy formula for LL AdS (\ref{F_LL}), that includes
backreaction of the gravitational fields, contains all thermodynamic
information about the holographic quantum system described by the partition
function $Z=e^{-F/T}$. The theory depends on a large number of parameters
(up to $[d/2]$ Lovelock gravity parameters, electromagnetic coupling
constants contained in the function $\mathcal{L}(F^{2})$ and the scalar
couplings in $\mathcal{F}(\Psi )$). Phase transitions will occur only for
some values of these parameters. In the present literature, their values are
chosen arbitrarily, in the points of the parameter space known that they
would have a phase transition.

We address a different question, focused on obtaining a criterion about the
parameters choice and, using it, a classification of all possible LL
superconductors. This would enable theoretical design\ of a superconductor
with desired features through a choice of the coupling constants. In that
way, a diversity of high-$T_{c}$ superconductors would be directly related
to a diversity of dual gravitational theories.

As the first step, we have to analyze the local and global minima of the
free energy using the renormalized formula (\ref{F_LL}) and detect all
possible phase transitions through the discontinuities in the free energy,
similarly to an analysis in the Landau-Ginzburg description of
superconductivity.

In a $d$-dimensional thermal QFT, the temperatures are high. Since the
gauge/gravity duality prescription relates $T$ to the Hawking temperature
proportional to the horizon area, high $T$ corresponds to large black hole
radius $r_{+}$. More precisely, in the gravity side, we have to require that
the black hole horizon is big, $r_{+}\gg \ell _{\text{eff}}$, that we can
also take as the first approximation in analytic calculations.

Thus, to obtain an initial information about possible phase transitions
--that is, to detect development of hair in the black hole due to presence
of the scalar field-- we need only an asymptotic solution, but calculated in
the subleading order. This enables to identify the point of the phase
transition and also calculate its critical exponent $\beta $ and the
critical temperature. Once there is a classification of the
superconductors, we can also calculate the transport coefficients
(superconductivity, energy gap, etc.), where further knowledge on the
behavior of the solution away from the asymptotic region is also required.

To be more precise, let us focus again to the EGB AdS gravity with the
matter. We have to power-expand in $\ell _{\text{eff}}/r<<1$ all unknown
functions $f(r),N(r),\Psi (r),\phi (r)$, and all given functions $\mathcal{F}%
(\Psi (r)),\mathcal{L(}F^{2}(r))$. Note that $\mathcal{F}$ and $\mathcal{L}$
can be chosen arbitrarily, as they define interactions between the gravity
and matter fields,\ whereas $f$, $N$, $\Psi $ and $\phi $ should be uniquely
determined for given boundary conditions set. Thus, assuming the known
asymptotic behavior, any function $X(r)$ can be expanded in powers of $1/r$,
where the constants $\ell _{\text{eff}}$ are absorbed in the coefficients $%
X_{n}$ for the sake of simplicity. We numerate\ only nonvanishing
coefficients $X_{n}$ with $n\geq 0$.

In general, we are interested in generic black holes, that is, the ones
where (\textit{i}) the equation of motion for the field $X(r)$ uniquely
determines the coefficient $X_{n}$ in the $n$-th order of the asymptotic
power expansion in terms of the leading order, and (\textit{ii}) there are
no anomalies, that is, no logarithmic terms in the expansion are present.

There will also be other particular solutions (that do not fulfill the
condition (\textit{i})) that have to be considered separately. The anomalies
(that do not satisfy (\textit{ii})) are of particular interest and they will
be addressed somewhere else.

For an asymptotically AdS spacetime that satisfies (\ref{AAdS planar}), we
seek for a solution in the form%
\begin{eqnarray}
f(r) &=&\frac{r^{2}}{\ell _{\text{eff}}^{2}}\left( 1+\frac{f_{0}}{r^{u}}+%
\frac{f_{1}}{r^{u_{1}}}+\mathcal{\cdots }\right) \,,  \notag \\
N(r) &=&1+\frac{N_{1}}{r^{s_{1}}}+\mathcal{\cdots }\text{\thinspace },
\notag \\
\Psi (r) &=&\frac{\Psi _{0}}{r^{\Delta }}+\frac{\Psi _{1}}{r^{\Delta _{1}}}%
+\cdots \,,  \notag \\
\phi (r) &=&\Phi +\frac{\phi _{0}}{r^{\lambda }}+\frac{\phi _{1}}{r^{\lambda
_{1}}}+\cdots \,,  \label{asymptotic}
\end{eqnarray}%
where $0<u<u_{n}<u_{n+1}$ and similarly for all other power factors. The
interaction is given in terms of a set of parameters that determine its
fall-off,
\begin{eqnarray}
\mathcal{F}(\Psi ) &=&\frac{\mathcal{F}_{0}}{r^{a}}+\frac{\mathcal{F}_{1}}{%
r^{a_{1}}}+\cdots \,,  \notag \\
\mathcal{L(}F^{2}) &=&\frac{\mathcal{L}_{0}}{r^{b}}+\frac{\mathcal{L}_{1}}{%
r^{b_{1}}}+\cdots \,.
\end{eqnarray}%
For example, $a=2\Delta $ implies that $\mathcal{F}$ behaves as the minimal
coupling of the scalar field. If $b=2\left( \lambda +1\right) $, the and NED
coupling belongs to a class of the Born-Infeld-like Lagrangians that have
the same weak-field behavior as the Maxwell electrodynamics (linear in $%
F^{2}=-2N\phi ^{\prime \,2}$)\textbf{.}

Solving the equations of motion at different orders of $1/r$ leads to
several branches of solutions, a couple of them contain the minimal scalar
coupling and Born-Infeld-like theories. A complete set of asymptotic
solutions will be discussed in detail in a classification of holographic
superconductors \cite{Aranguiz-Miskovic}.

Let us mention some common features of all solutions, mostly independent on
the dynamics. The scalar field behaves as $\Psi \sim r^{-\left( d-\Delta
\right) }\,\Psi _{\text{source}}$ for large $r$. A dual operator $\mathcal{O}%
_{\Psi }$ coupled to the scalar in a holographic CFT$_{d}$ has, therefore,
the conformal dimension $\dim \mathcal{O}_{\Psi }=\Delta $, that must be
greater than the CFT unitarity bound, $\dim O_{\Psi }\geq \frac{d-2}{2}$\
\cite{Minwalla:1997ka}. The equation of motion (\ref{KG_eval}) gives a
well-known relation that determines $\Delta $ in terms of the scalar mass,%
\begin{equation}
\Delta \left( \Delta -d\right) =m^{2}\ell _{\text{eff}}^{2}\,\,,
\end{equation}%
where it was assumed that there is no conformal anomaly, that is, $\Delta
\neq d$, and the electromagnetic interaction does not modify the asymptotic
sector, $a>2\Delta -2$. This equation leads to two branches for the scalar
field, $\Delta =\Delta _{+}$ and $\Delta =\Delta _{-}=d-\Delta _{+}$. If $%
d-\Delta <0$, the scalar field is divergent in IR sector and the dual
operator with $\dim \mathcal{O}_{\Psi }>d$ is irrelevant deformation of the
theory. If the scalar field falls off sufficiently fast ($d-\Delta \geq 0$),
$\mathcal{O}_{\Psi }$ is relevant or marginal operator satisfying $\dim
\mathcal{O}_{\Psi }\leq d$, and it can be turn on without destroying the UV
fixed point of the dual QFT$_{d}$. We are interested in the last case
because we want to interpret $\Psi (r)$ as the black hole hair, so we need
it to be regular everywhere, including the IR region. These conditions allow
for a tachyonic scalar field that still gives a unitary CFT, known as the
Breitenlohner-Freedman window\textbf{\ }\cite%
{Breitenlohner:1982bm,Breitenlohner:1982jf}. As respect to a scalar
observable in QFT$_{d}$ that will play a role of the order parameter of a
superconductor, it can be only a normalizable mode. It is common to take the
source$\ $to be switched off, $\Psi _{\text{source}}=0$, and then $\Psi
_{0}=\left\langle \mathcal{O}_{\Psi }\right\rangle $ is a normalizable
operator \cite{Herzog:2010vz,Hartnoll:2009sz} and it is the leading order in
the near-boundary expansion, as presented in (\ref{asymptotic}).

The first integral of the NED equation (\ref{Maxwell_t}) gives an
integration constant that is related to the Noether charge,%
\begin{equation}
4r^{d-1}\sqrt{N}E\,\dfrac{d\mathcal{L}}{dF^{2}}=-\frac{2\kappa ^{2}Q}{V_{d-1}%
}+\int dr\,\frac{r^{d-1}\mathcal{F}\phi }{\sqrt{N}f}\,.
\end{equation}%
Non-vanishing $Q$ fixes the fall-off of the electric potential as $\lambda
=b-d$. The source $\phi _{0}$ is, then, an algebraic function of the
electric charge, $Q=Q(\phi _{0})$, given by%
\begin{equation}
\frac{\mathcal{L}_{0}}{\phi _{0}}=\frac{2\kappa ^{2}(b-d)(b-d+1)}{bV_{d-1}}%
\,Q\,.
\end{equation}

Similarly, the first integral of the gravitational equation $\mathcal{E}%
_{t}^{t}$ gives rise to an integration constant that depends on the black
hole mass obtained from the\ Noether formula (\ref{M}),
\begin{equation}
r^{d-2}Nf-\alpha \,\left( d-2\right) \left( d-3\right) r^{d-4}N^{2}f^{2}-%
\frac{r^{d}}{\ell ^{2}}=-\frac{2\kappa ^{2}M}{\left( d-1\right) V_{d-1}}%
+\int dr\,\dfrac{2r^{d-1}}{d-1}\,T_{t}^{t}(r)\,.  \label{Grav_tt_Int}
\end{equation}%
The requirement $M\neq 0$ determines the leading order of this equation as $%
u=d$ and also imposes that the subleading order of $N(r)$ must be small
enough, $s_{1}>d$. The integration constant is then related to the
coefficient $f_{0}$ by%
\begin{equation}
f_{0}=-\frac{2\kappa ^{2}\ell _{\text{eff}}^{2}\,}{\left( d-1\right)
V_{d-1}\left( 1-\frac{2\alpha }{\ell _{\text{eff}}^{2}}\left( d-2\right)
\left( d-3\right) \right) }\,M\,.
\end{equation}%
It is worthwhile noticing that this is not the unique way to obtain $M\neq 0$
from the asymptotic expansion -- other branches can involve the scalar field
contribution, and even the electric charge. However, it remains to analyse
how these new branches modify the UV sector of a dual QFT.

We shall not write here the subleading orders of the solutions. Each order
of the field equations solves one of the coefficients $X_{n}$ as polynomials
in $f_{0}$, $\phi _{0}$ and $\psi _{0}$, that is, $M$, $Q$ and $\left\langle
\mathcal{O}_{\Psi }\right\rangle $. Furthermore, imposing the boundary
conditions on the horizon will involve also the parameters $T$ and $r_{+}$,
and eliminate the mass as an independent variable. As a result, using the
asymptotic solution, the free energy $F(T,Q,\left\langle \mathcal{O}_{\Psi
}\right\rangle )$ can be cast into the form%
\begin{equation}
\tilde{F}(r_{+},\phi _{0},\Psi _{0})=F(T(r_{+},\phi _{0},\Psi _{0}),Q(\phi
_{0}),\Psi _{0})\,.
\end{equation}%
An analysis of the extrema of this function for constant $Q$ and $T$ is then
a well-posed problem.

\section{Conclusions}

Motivated by an application of AdS/CFT correspondence to $d$-dimensional
high-T$_{c}$ superconductors that do not have a generally accepted
theoretical model, we study the St\"{u}ckelberg scalar field in ($d+1$%
)-dimensional asymptotically AdS spacetime coupled to gravitational and
electromagnetic fields. In the AdS$_{d+1}$ gravity side, the black hole
solution is associated to a thermal, dual QFT$_{d}$ and the scalar field
couples to the order parameter of the superconductor in QFT$_{d}$.

In order to involve a wider class of holographic superconductors, we include
non-linear effects in the gravity action: the GB term and LL
generalization of General Relativity that is polynomial in the Riemann
curvature and has $[d/2]$ coupling constants, NED field described by an
arbitrary function $\mathcal{L}(F^{2})$ and the St\"{u}ckelberg modification
of the scalar field kinetic term through the function $\mathcal{F}(\Psi )$.
Of course, because we are ultimately interested in lower-dimensional QFTs,
one should also introduce other higher-order curvature terms not of Lovelock type.

On the other hand, study of phase transitions in QFTs can teach us about instabilities of black hole solutions in LL-AdS gravities, as well.

We focus to the maximally symmetric, charged, AdS black holes with flat
horizons. Using the Kounterterm regularization of the AdS gravity action
that is universal for all LL gravities, we obtain the exact formula for a IR
divergence-free Euclidean action and the finite Noether charges. Depending
on the choice of the NED boundary term, the Euclidean action is identified
with the appropriate thermodynamic potential in canonical or grand canonical
ensemble. These thermodynamic potentials, obtained from the gravitational
quantum statistical relation, correctly reproduce the Legendre
transformation of the internal energy of the superconductor in a dual QFT.

We also note that the effect of backreaction cancels all divergences in the
scalar field. Furthermore, the St\"{u}ckelberg scalar does not contribute to
the black hole mass. As regards to the entropy, it still respects the
horizon area law because, in the flat transversal section, there is no the
LL contribution to the Euclidean action, and the St\"{u}ckelberg field does
not induce extra terms that were noted for other kind of non-minimal
coupling of the scalar field.

Let us emphasize that a novelty of our analytic method is that it provides
a holographic free energy formula for the St\"{u}ckelberg superconductor without
using any approximation. The phase transitions now can be analyzed as in
the Ginzburg-Landau model, that is, by studying its extrema. This technique
can also be applied to 5D Einstein-Hilbert AdS gravity. However,
thanks to a regularization method employed, the formula is extended to all
LL-AdS gravities in higher dimensions.

A natural direction for future research is
to discuss how these results can be used to obtain physical information
about a holographic superconductor. More explicitly, the study of local and
global minima of the free energy of a holographic superconductor can
identify all possible discontinuities of second order associated to phase
transitions. This method would enable to classify all LL AdS superconductors
in the space of parameters. By looking only at the asymptotic solutions in
the leading and subleading orders, it is possible to obtain information
about the critical temperature and critical exponent. As concluding remarks,
we discuss about general features of the asymptotic solutions.

\section*{Acknowledgment}

The authors thank Rodrigo Olea and Dumitru Astefanesei for useful comments.
L.A. would also like to thank Stefan Theisen for hospitality during her stay
at Max Planck Institute for Gravitational Physics (Albert Einstein
Institute). This work was supported by Chilean FONDECYT grant Nb.1110102.
O.M. is grateful to DII-PUCV for their support through the
project Nb.123.711/2011. The work of L.A. is financed in part by Chilean grants
NAC-\emph{doctorado} Nb.21090754, BCH-\emph{pasantia doctoral} Nb.75120006,  and the UTFSM projects PIIC/2011-2013.

\appendix{}

\section{Black hole ansatz identities \label{TBH}}

The metric of the maximally symmetric, planar black hole in a spacetime with
local coordinates $x^{\mu }=(t,r,y^{m})$ has the form%
\begin{equation}
g_{\mu \nu }=\left(
\begin{array}{ccc}
-f(r) & 0 & 0 \\
0 & \frac{1}{f(r)N(r)} & 0 \\
0 & 0 & r^{2}\delta _{mn}%
\end{array}%
\right) \,,\qquad \sqrt{-g}=\frac{r^{d-1}}{\sqrt{N}}\,,
\end{equation}%
where $f(r)\geq 0$ and $N(r)>0$. \ In this ansatz, the Riemann tensor $%
R_{\lambda \rho }^{\mu \nu }=R_{\ \sigma \lambda \rho }^{\mu }g^{\sigma \nu
} $ has non-vanishing components
\begin{equation}
\begin{array}{ll}
R_{tr}^{tr}=-\dfrac{1}{2}\,\left( Nf^{\prime \prime }+\dfrac{1}{2}%
\,f^{\prime }N^{\prime }\right) \,,\medskip \qquad & R_{tm}^{tn}=-\dfrac{%
Nf^{\prime }}{2r}\,\delta _{m}^{n}\,, \\
R_{m_{1}m_{2}}^{n_{1}n_{2}}=-\dfrac{fN}{r^{2}}\,\delta
_{m_{1}m_{2}}^{n_{1}n_{2}}\,, & R_{rm}^{rn}=-\dfrac{\left( fN\right)
^{\prime }}{2r}\,\delta _{m}^{n}\,,%
\end{array}%
\end{equation}%
plus the components obtained from the antisymmetry in the pairs of indices.
The prime denotes radial derivative.

The components of the Ricci tensor, $R_{\nu }^{\mu }=R_{\nu \lambda }^{\mu
\lambda }$, are given by%
\begin{eqnarray}
R_{t}^{t} &=&-\dfrac{1}{2r}\left[ rNf^{\prime \prime }+\frac{1}{2}%
\,rf^{\prime }N^{\prime }+\left( d-1\right) Nf^{\prime }\right] \,,  \notag
\\
R_{r}^{r} &=&-\dfrac{1}{2r}\left[ rNf^{\prime \prime }+\frac{1}{2}%
\,rf^{\prime }N^{\prime }+\left( d-1\right) \left( fN\right) ^{\prime }%
\right] \,,  \notag \\
R_{m}^{n} &=&-\dfrac{1}{r^{2}}\,\delta _{m}^{n}\left[ rNf^{\prime }+\frac{1}{%
2}\,rfN^{\prime }+\left( d-2\right) fN\right] \,.
\end{eqnarray}

The Ricci scalar, $R=R_{\mu }^{\mu }$, then reads%
\begin{equation}
R=-\dfrac{1}{r^{2}}\,N\left[ r^{2}f^{\prime \prime }+2\left( d-1\right)
rf^{\prime }+\left( d-1\right) \left( d-2\right) f\right] -\frac{1}{2r}%
\,N^{\prime }\left[ rf^{\prime }+2\left( d-1\right) f\right] \,,
\end{equation}%
and the GB term is%
\begin{eqnarray}
\frac{1}{4}\,\delta _{\nu _{1}\cdots \nu _{4}}^{\mu _{1}\cdots \mu
_{4}}\,R_{\mu _{1}\mu _{2}}^{\nu _{1}\nu _{2}}R_{\mu _{3}\mu _{4}}^{\nu
_{3}\nu _{4}} &=&\frac{2\left( d-1\right) \left( d-2\right) }{r^{4}}\,\left[
NN^{\prime }\left( \left( d-3\right) rf^{2}+\dfrac{3}{2}\,r^{2}ff^{\prime
}\right) +\right.  \notag \\
&&+\left. N^{2}\left( r^{2}ff^{\prime \prime }+r^{2}f^{\prime 2}+2\left(
d-3\right) rff^{\prime }+\frac{1}{2}\,\left( d-3\right) \left( d-4\right)
f^{2}\right) \right] .  \label{GB_ansatz}
\end{eqnarray}

The components of the Einstein tensor with the negative cosmological
constant have the form%
\begin{eqnarray}
G_{r}^{r} &=&\dfrac{d-1}{2r^{2}}\left[ rNf^{\prime }+\left( d-2\right) Nf-d\,%
\frac{r^{2}}{\ell ^{2}}\right] \,,  \notag \\
G_{t}^{t} &=&G_{r}^{r}+\dfrac{d-1}{2r}\,N^{\prime }f\,,  \notag \\
G_{m}^{n} &=&\dfrac{1}{2r^{2}}\,\delta _{m}^{n}\left[ r^{2}Nf^{\prime \prime
}+2\left( d-2\right) rNf^{\prime }+\frac{1}{2}\,r^{2}N^{\prime }f^{\prime
}\right.  \notag \\
&&+\left. \left( d-2\right) rN^{\prime }f+\left( d-2\right) \left(
d-3\right) Nf-d\left( d-1\right) \frac{r^{2}}{\ell ^{2}}\right] \,,
\end{eqnarray}%
and the Lanczos tensor\ in components reads%
\begin{eqnarray}
H_{r}^{r} &=&-\alpha \,\left( d-1\right) \left( d-2\right) \left( d-3\right)
\,\frac{f}{r^{3}}\,N^{2}\left( f^{\prime }+\frac{d-4}{2r}\,f\right) \,,
\notag \\
H_{t}^{t} &=&H_{r}^{r}-\alpha \,\left( d-1\right) \left( d-2\right) \left(
d-3\right) \,\frac{f^{2}NN^{\prime }}{r^{3}}\,,  \notag \\
H_{m}^{n} &=&-\alpha \left( d-2\right) \left( d-3\right) \,\delta _{m}^{n}\,%
\dfrac{1}{r^{2}}\left[ N^{2}ff^{\prime \prime }+N^{2}f^{\prime 2}+\dfrac{3}{2%
}\,NN^{\prime }ff^{\prime }\right.  \notag \\
&&+\left. \dfrac{1}{r}\left( d-4\right) \left( 2N^{2}ff^{\prime }+NN^{\prime
}f^{2}\right) +\left( d-4\right) \left( d-5\right) \,\dfrac{f^{2}N^{2}}{%
2r^{2}}\right] \,.
\end{eqnarray}

We also write the following auxiliary expressions,%
\begin{eqnarray}
G_{m}^{m} &=&\frac{\left( r^{d-1}G_{r}^{r}\right) ^{\prime }}{r^{d-2}}-%
\dfrac{d-1}{4}\,N^{\prime }f^{\prime }\,,  \notag \\
H_{m}^{m} &=&\frac{\left( r^{d-1}H_{r}^{r}\right) ^{\prime }}{r^{d-2}}%
+\alpha \left( d-1\right) \left( d-2\right) \left( d-3\right) \,\dfrac{%
NN^{\prime }ff^{\prime }}{2r^{2}}\,.  \label{Hmm}
\end{eqnarray}

\section{Ordinary hypergeometric function \label{Hyper}}

The ordinary hypergeometric function in the integral representation is given
by
\begin{equation}
\left. _{2}F_{1}\right. \left( a,b;c;z\right) =\frac{\Gamma (c)}{\Gamma
(b)\Gamma (c-b)}\int\limits_{0}^{1}dt\,\frac{t^{b-1}\left( 1-t\right)
^{c-b-1}}{\left( 1-zt\right) ^{a}}\,,\qquad \Re (c)>\Re (b)>0\,.
\end{equation}%
It can be expanded in the hypergeometric series whose coefficients are given
by the Pochhammer symbol, $(a)_{p}$ ,
\begin{eqnarray}
\left. _{2}F_{1}\right. \left( a,b;c;z\right) &=&1+\frac{ab}{c}\,z+\frac{%
a\left( a+1\right) b\left( b+1\right) }{2c\left( c+1\right) }\,z^{2}+\cdots
\notag \\
&=&\sum\limits_{p=0}^{\infty }\frac{(a)_{p}(b)_{p}}{(c)_{p}}\,\frac{z^{p}}{%
p!}\,,
\end{eqnarray}%
that converges when $c$ is not a negative integer for all $|z|<1$, and on
the unit circle $|z|=1$ if $\Re (c-a-b)>0$.

In this text, we need the following integrals represented in terms of the
hypergeometric functions,%
\begin{eqnarray}
\int\limits_{0}^{1}du\,\left( -\beta +u^{2}w\,\right) ^{n-1} &=&\left(
-\beta \right) ^{n-1}\left. _{2}F_{1}\right. \left( 1-n,\frac{1}{2};\frac{3}{%
2};\frac{w}{\beta }\right) \,,  \notag \\
\int\limits_{0}^{1}ds\,\left( 1-s^{2}\right) ^{n-1} &=&\left.
_{2}F_{1}\right. \left( 1-n,\frac{1}{2};\frac{3}{2};1\right) =\frac{%
2^{2n-2}\left( n-1\right) !^{2}}{\left( 2n-1\right) !}\,,
\end{eqnarray}%
as well as the integrals used in Sections \ref{Var} and \ref{Con},
\begin{eqnarray}
\mathcal{I}_{n}(\beta ,w) &=&\int\limits_{0}^{1}du\int\limits_{0}^{u}ds\,%
\left[ -u^{2}\beta +\left( 2n-1\right) s^{2}w\right] \left( -u^{2}\beta
+s^{2}w\right) ^{n-2}=\frac{\left( w-\beta \right) ^{n-1}}{2n}\,,  \notag \\
\mathcal{J}_{n}(\beta ,w) &=&\int\limits_{0}^{1}du\int\limits_{0}^{u}ds\,%
\left[ -\left( 2n-1\right) u^{2}\beta +s^{2}w\right] \left( -u^{2}\beta
+s^{2}w\right) ^{n-2}  \notag \\
&=&\left( -\beta \right) ^{n-1}\left. _{2}F_{1}\right. \left( 1-n,\frac{1}{2}%
;\frac{3}{2};\frac{w}{\beta }\right) -\frac{\left( w-\beta \right) ^{n-1}}{2n%
}\,.
\end{eqnarray}

\end{document}